\documentclass[conference]{IEEEtran}
\IEEEoverridecommandlockouts
\usepackage{cite}
\usepackage{amsmath,amssymb,amsfonts,braket}
\usepackage{algorithmicx}
\usepackage{graphicx}
\usepackage{textcomp}
\usepackage{xcolor}
\usepackage{algorithm,balance}
\usepackage[noend]{algpseudocode}

\usepackage{subfigure}

\usepackage{epsfig}
\usepackage{algorithm}

\newcommand{\remove}[1]{}

\newtheorem{lemma}{Lemma}
\newtheorem{corollary}{Corollary}

\newcommand{\ls}[1]
   {\dimen0=\fontdimen6\the\font
    \lineskip=#1\dimen0
    \advance\lineskip.5\fontdimen5\the\font
    \advance\lineskip-\dimen0
    \lineskiplimit=.9\lineskip
    \baselineskip=\lineskip
    \advance\baselineskip\dimen0
    \normallineskip\lineskip
    \normallineskiplimit\lineskiplimit
    \normalbaselineskip\baselineskip
    \ignorespaces
   }

\newcommand{\singlefig}[3]{
\begin{figure}
\centerline{
    \setlength{\epsfysize}{0.28\textwidth}
 \epsffile{\Figdir#1}
} \caption{#2} \label{fig:#3}
\end{figure}
}

\newcommand{\tydubfigsingle}[6]{
\begin{figure}
\centerline{
    \begin{minipage}{0.30\textwidth}
      \begin{center}
        \leavevmode
        \setlength{\epsfxsize}{0.85\textwidth}
        \setlength{\epsfysize}{0.72\textwidth}
        \epsffile{\Figdir#1}
       \newline{\small (a) #2}
      \end{center}
    \end{minipage}
    \hspace*{-1cm}
    \begin{minipage}{0.30\textwidth}
      \begin{center}
        \leavevmode
        \setlength{\epsfxsize}{0.85\textwidth}
        \setlength{\epsfysize}{0.72\textwidth}
        \epsffile{\Figdir#3}
       \newline{\small (b) #4}
      \end{center}
    \end{minipage}
} \caption{#5}\label{fig:#6}
\end{figure}
}

\newcommand{\triplefig}[8]{
\begin{figure*}[t]
\centerline{
    \begin{minipage}{0.32\textwidth}
      \begin{center}
        \leavevmode
        \setlength{\epsfxsize}{0.85\textwidth}
        \setlength{\epsfysize}{0.85\textwidth}
        \epsffile{\Figdir#1}\\
       {\small (a) #2}
      \end{center}
    \end{minipage}
    \begin{minipage}{0.32\textwidth}
      \begin{center}
        \leavevmode
        \setlength{\epsfxsize}{0.85\textwidth}
        \setlength{\epsfysize}{0.85\textwidth}
        \epsffile{\Figdir#3}\\
       {\small (b) #4}
      \end{center}
    \end{minipage}
    \begin{minipage}{0.32\textwidth}
      \begin{center}
        \leavevmode
        \setlength{\epsfxsize}{0.85\textwidth}
        \setlength{\epsfysize}{0.85\textwidth}
         \epsffile{\Figdir#5}\\
       {\small (c) #6}
      \end{center}
    \end{minipage}
} \caption{#7}\label{fig:#8}
\end{figure*}
}

\newcommand{\triplefigbig}[8]{
\begin{figure*}[t]
\centerline{
    \begin{minipage}{0.30\textwidth}
      \begin{center}
        \leavevmode
        \setlength{\epsfxsize}{0.85\textwidth}
        \setlength{\epsfxsize}{\textwidth}
        \setlength{\epsfysize}{0.85\textwidth}
        \setlength{\epsfysize}{\textwidth}
        \epsffile{\Figdir#1}\\
       {\small (a) #2}
      \end{center}
    \end{minipage}
    \hspace{0.2in}
    \begin{minipage}{0.30\textwidth}
      \begin{center}
        \leavevmode
        \setlength{\epsfxsize}{0.85\textwidth}
        \setlength{\epsfxsize}{\textwidth}
        \setlength{\epsfysize}{0.85\textwidth}
        \setlength{\epsfysize}{\textwidth}
        \epsffile{\Figdir#3}\\
       {\small (b) #4}
      \end{center}
    \end{minipage}
    \hspace{0.2in}
    \begin{minipage}{0.30\textwidth}
      \begin{center}
        \leavevmode
        \setlength{\epsfxsize}{0.85\textwidth}
        \setlength{\epsfxsize}{\textwidth}
        \setlength{\epsfysize}{0.85\textwidth}
        \setlength{\epsfysize}{\textwidth}
         \epsffile{\Figdir#5}\\
       {\small (c) #6}
      \end{center}
    \end{minipage}
} \caption{#7}\label{fig:#8}
\end{figure*}
}

\newcommand{\Figdir}{./}

\newcommand{\kb}[1]{\ket{#1}\bra{#1}}

\newcommand\note[3]{{\textcolor{#1}{[\textsf{#2}: #3]}}}
\newcommand{\WK}[1]{\note{blue}{WK}{#1}}
\newcommand{\OA}[1]{\note{orange}{OA}{#1}}

\newcommand{\rawkey}{\mathcal{RK}}
\newcommand{\secretkey}{\mathcal{K}}

\algnewcommand\algorithmicforeach{\textbf{for each}}
\algdef{S}[FOR]{ForEach}[1]{\algorithmicforeach\ #1\ \algorithmicdo}

\def\BibTeX{{\rm B\kern-.05em{\sc i\kern-.025em b}\kern-.08em
    T\kern-.1667em\lower.7ex\hbox{E}\kern-.125emX}}
\begin{document}

\title{Efficient Routing for Quantum Key Distribution Networks\\
}

\author{
\IEEEauthorblockN{Omar Amer}
\IEEEauthorblockA{\textit{Computer Science Department} \\
\textit{University of Connecticut}\\
Storrs CT, USA \\
omar.amer@uconn.edu}
\and
\IEEEauthorblockN{Walter O. Krawec}
\IEEEauthorblockA{\textit{Computer Science Department} \\
\textit{University of Connecticut}\\
Storrs CT, USA\\
walter.krawec@uconn.edu}
\and
\IEEEauthorblockN{Bing Wang}
\IEEEauthorblockA{\textit{Computer Science Department} \\
\textit{University of Connecticut}\\
Storrs CT, USA \\
bing@uconn.edu}
}

\maketitle

\begin{abstract}
    As quantum key distribution becomes increasingly practical, questions of how to effectively employ it in large-scale networks and over large distances becomes increasingly important. To that end, in this work, we model the performance of the E91 entanglement based QKD protocol when operating in a network consisting of both quantum repeaters and trusted nodes. We propose a number of routing protocols for this network and compare their performance under different usage scenarios. Through our modeling, we investigate optimal placement and number of trusted nodes versus repeaters depending on device performance (e.g., quality of the repeater's measurement devices). Along the way we discover interesting lessons determining what are the important physical aspects to improve for upcoming quantum networks in order to improve secure communication rates.
\end{abstract}
\section{Introduction}

Quantum key distribution (QKD) allows for the establishment of information theoretically secure secret keys between two or more parties.  However, despite their great potential, these systems face several critical shortcomings when attempting to implement them in practice.  Of particular importance are improving the speed and distance of these systems.  To overcome these limitations, quantum networks are often used, consisting of \emph{trusted nodes} and, in the near future, \emph{quantum repeaters} \cite{repeater1,repeater2}.  Since quantum repeaters are still a developing technology, current QKD networks established in various metropolitan areas consist only of end-users and trusted nodes \cite{elliott2004darpa,chen2010metropolitan,peev2009secoqc,zhang2018large,sasaki2011field}.  However, progress in developing stable quantum repeaters has been accelerating of late and so it is vital to begin developing suitable routing algorithms for networks consisting of a mixture of both repeaters and trusted nodes.  Indeed, developing efficient routing algorithms to operate in this setting is vital to the future performance of such networks.

Numerous work has been done investigating the performance of quantum networks consisting only of quantum repeaters (in addition to end users) \cite{van2013path,azuma2016fundamental,hahn2019quantum,vardoyan2019stochastic,pirandola2016capacities,wallnofer2019multipartite,pirandola2019end,li2020effective}, including the development and analysis of new routing protocols specific to that scenario \cite{caleffi2017optimal,gyongyosi2017entanglement,sazim2015retrieving,pant2019routing,chakraborty2019distributed}.  Mostly, the goal of such networks is to establish end-to-end entanglement between end users (which may, then, be used for QKD for instance).  However, QKD is a more practical technology today and there are several methods to improve their performance, such as through trusted nodes.  Thus it is also important to study QKD-specific networks.  Such work has been done investigating practical QKD-specific networks consisting predominately of trusted nodes and end users \cite{dianati2008architecture,qkd-network-stack,yang2017qkd}.  However, as repeater technology progresses, the desire to move away from trusted node technology, and their inherent security concerns, will become stronger.

Thus, it is important to begin investigating near-future QKD-specific networks consisting predominately of quantum repeaters, however with a minority of trusted nodes.  Furthermore, these networks will allow for multiple paths to be established between both repeaters and trusted nodes (i.e., repeaters are not used only to extend the distance between trusted nodes, but will be an integral part of the network interior).  These are the networks we consider in this paper.  In particular, we consider a grid topology in which it is possible to establish multiple paths between Alice, Bob, and various trusted nodes though a complex network of repeaters. Our goal is to understand what routing protocols can lead to efficient key distribution rates between end users and to understand how the quality of the repeaters and the number of trusted nodes, affects the performance of this network.

In this paper, we devise and evaluate three different routing algorithms for this QKD network: one requires global state information and the other two are decentralized, distributed, algorithms requiring only local state information.
We perform a rigorous evaluation, through simulations, of the performance of our algorithms in a variety of scenarios considering network size; quantum repeater quality; distance between nodes; quantity and specific location of trusted nodes; and channel noise.  Our results show that the careful design of routing algorithms is vital to establishing efficient key-distribution rates between users.  Along the way, we discover several fascinating properties of these networks which may be of great importance to operators of this future network architecture.

\subsection{Quantum Key Distribution}
Quantum key distribution (QKD) protocols operate in two stages: In the \emph{quantum communication stage}, users Alice and Bob use the quantum channel and the authenticated classical channel to attempt to establish a \emph{raw key} which is a classical bit string that is partially correlated (there may be errors in the quantum channel leading to errors in the raw key) and partially secret (an adversary may have some information on the raw key).  Since this raw key cannot be used directly for cryptographic purposes, a second, purely classical, stage is performed, running an error correction protocol (leaking additional information to an adversary) followed by privacy amplification, which leads to the final secret key.  An important metric in any QKD protocol is the protocol's \emph{key-rate} defined to be the ratio of secret key bits to the size of the raw key.  For a general survey of QKD protocols, the reader is referred to \cite{QKD-survey-old,QKD-Survey}.

In this work, we are interested in the theoretical performance of a network of QKD systems.  As such, we consider ideal single-qubit sources (though potentially there is loss due to the fiber channel connecting nodes).  Furthermore, we do not consider finite key effects \cite{scarani2008quantum} or imperfect sampling, in an attempt to understand the theoretical behavior of the network.  Under this setting, the key-rate of the E91 \cite{QKD-E91} protocol, which we adopt as the QKD protocol used in our network, is found to be \cite{QKD-BB84-rate1,QKD-renner-keyrate}: $\texttt{rate} = 1-2h(Q)$, where $Q$ is the bit error rate in the raw key and $h(\cdot)$ is the binary entropy function.  Note that, if we did not assume single-qubit sources and, instead, weak coherent sources, we would need to use an alternative key-rate expression or, perhaps, the decoy state BB84 protocol \cite{lo2005decoy}; these are interesting issues we leave as future work.

Point-to-point QKD systems are often implemented over fiber channels (though free-space operation is also possible if one has direct line of sight).  One limitation to QKD protocols is their intolerance to qubit loss \cite{pirandola2017fundamental,QKD-Survey}; since loss over a fiber channel scales exponentially with distance, this is particularly problematic for ground operation.  To overcome this, one may use \emph{quantum repeaters} \cite{repeater1,repeater2} and \emph{trusted nodes}.  Both systems are placed between users, thus halving the total distance a qubit is required to travel; furthermore, multiple such nodes may be chained together thus further decreasing the distance.  Quantum repeaters utilize entanglement swapping to produce a shared entangled bit between two users.  Trusted nodes act as users in a QKD protocol, establishing a key $k_{AT}$ with Alice and a second, independent key, $k_{TB}$ with Bob.  The trusted node then sends to $B$ (using an authenticated classical channel) the value $k_{AT}\oplus k_{TB}$, where $\oplus$ is the bit-wise XOR.  This allows Alice and Bob to share a classical key, though the trusted node also shares this key.

The advantage to quantum repeaters is that the final key $A$ and $B$ produce is independent of the repeater's knowledge.  That is, if a repeater is controlled by an adversary, it cannot learn the final secret key.  However, the technology for repeaters requires short-term quantum memories which is a difficult engineering challenge (though the technology is rapidly advancing).  Trusted nodes have the advantage of simplicity as they are no different in technology than Alice or Bob.  However, they must remain trusted and safe from an adversary as they do have full knowledge of the secret key.


\subsection{Related Work}
There has been recent research in analyzing routing protocols, along with the general behavior and performance, of quantum networks consisting of end-users and quantum repeaters, but no trusted nodes; see for instance \cite{van2013path,azuma2016fundamental,caleffi2017optimal,gyongyosi2017entanglement,hahn2019quantum,pant2019routing,vardoyan2019stochastic,chakraborty2019distributed,pirandola2016capacities,wallnofer2019multipartite,pirandola2019end} with \cite{caleffi2017optimal,gyongyosi2017entanglement,pant2019routing,chakraborty2019distributed} giving particular focus on specific routing protocols for certain network topologies in generating shared entanglement between users, which is a stronger resource than QKD. However, QKD is, currently, a more practical and mature technology and, furthermore, there are numerous methods of improving the performance (speed and distance) of these systems that are unavailable to entanglement generation networks, namely the use of trusted nodes.  Thus, while routing protocols designed for an entanglement-generation quantum internet can be used also for QKD, more efficient systems may exist and, furthermore, solutions incorporating both trusted nodes and repeaters are vital for near-term deployment of this technology and are the networks we consider in this work.

Early work on QKD-specific networks focused on algorithms for optical-based switching networks \cite{toliver2003experimental}.  Such networks were used in practice, for instance the 2004 DARPA network in the Boston area \cite{elliott2004darpa,elliott2005current} and the 2009 network in Hefei \cite{chen2010metropolitan}, utilized such an architecture (the Hefei network also utilized trusted nodes).  These networks consisted of end-users and optical switches allowing users to route qubits to each other.

A more capable QKD network consists of end-users and trusted nodes and is the most common of network architectures for practical QKD networks in operation today such as the SECOQC network in Vienna \cite{peev2009secoqc}, the previously mentioned Hefei network \cite{chen2010metropolitan}, the Tokyo QKD network \cite{sasaki2011field} and the Beijing-Shanghai network \cite{zhang2018large}.  In terms of routing algorithms, the Vienna network used a variant of the OSPF routing algorithm \cite{dianati2008architecture}.  In \cite{tanizawa2016routing}, required functionalities of a QKD network routing algorithm were described for this form of network and a routing protocol was proposed, while \cite{yang2017qkd} began constructing routing protocols for larger scale networks evaluated through simulations. A novel quality of service model for these networks, and new routing protocols, were introduced in \cite{mehic2019novel}.  A complete QKD network stack for such networks was proposed in \cite{qkd-network-stack}, along with advanced routing functionalities through trusted nodes taking into account channel performance and key-rate demands, while in \cite{li2020mathematical}, a mathematical model was proposed for analyzing such trusted-node QKD networks.

The severe limitation to trusted nodes is that, as their name implies, they must be trusted.  Indeed, a trusted node is fully aware of the secret key which $A$ and $B$ distill (multi-path routing is one counter-measure to this, though still imperfect).  An advancement from this would be the use of quantum repeaters.  However, while the technology for these is advancing rapidly, it is more likely that in the near-future, any QKD network will consist of both repeaters and trusted nodes.  Thus, developing new and efficient routing algorithms specific for this network technology, is vital.  In this work, we consider networks that consist of a majority of repeaters with a minority of trusted nodes, attempting to discover optimal routing protocols for this scenario and to understand the behavior of such a network.  Note that in \cite{qkd-network-stack}, the use of quantum repeaters in a chain to link together different trusted nodes was considered an option; here, however, we are considering more complex networks of repeaters, allowing for potential multi-path routing options between both trusted nodes and multiple repeaters. 

\section{Model and Simulator}\label{sec:model}

In this work we consider a grid network consisting of $N\times N$ nodes with each node connected to at most four others, their immediate neighbors up, down, left, and right.  These connections represent fiber links allowing for the direct transmission of qubits between neighboring nodes only.  We assume a classical communication network allowing any pair of nodes to send classical messages.  We do not assume this communication is secret; however for all user and trusted node communication we do assume it is authenticated.  Each node in this network may be either a User (Alice or Bob); a Trusted Node; or a Quantum Repeater.  We place Alice at the lower-left corner of the network and Bob in the upper-right (thus, these users have only two neighbors they are connected to).  Later, we will simulate alternative numbers and locations of the trusted nodes; however, predominately the other nodes will consist of quantum repeaters in contrast to current day QKD networks.

\begin{figure}
    \centering
    \includegraphics{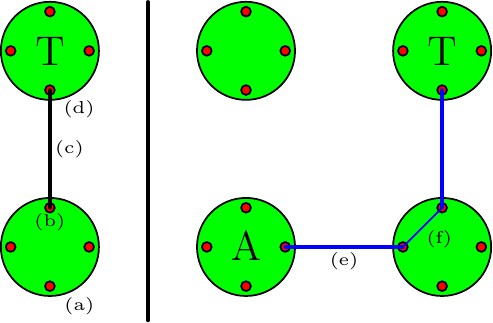}
    \caption{A legend for our network diagrams. Left: (a) denotes a node in the network, specifically a repeater with 4 (b) connection terminals from which entangled photons are sent and received over (c) fiber optic connections between other repeaters and (d) trusted nodes. Right: an entangled channel (e) connecting multiple trusted nodes through intra-node entanglement (f).}
    \label{fig:guide}
\end{figure}

\triplefig{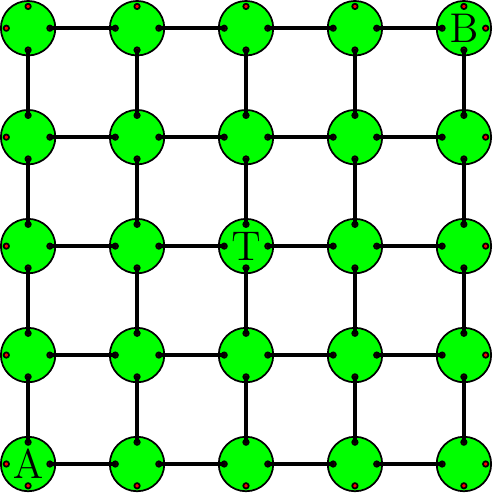}{Initial network}{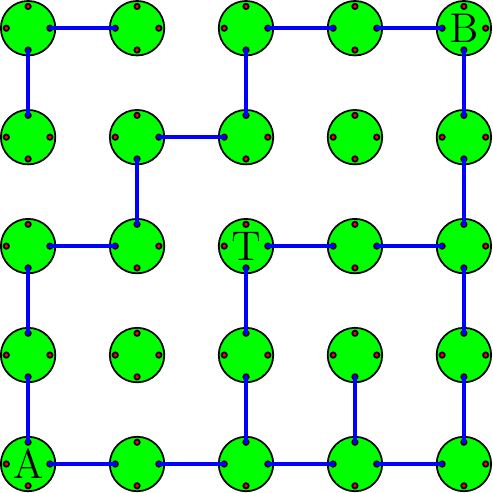}{Stage 1}{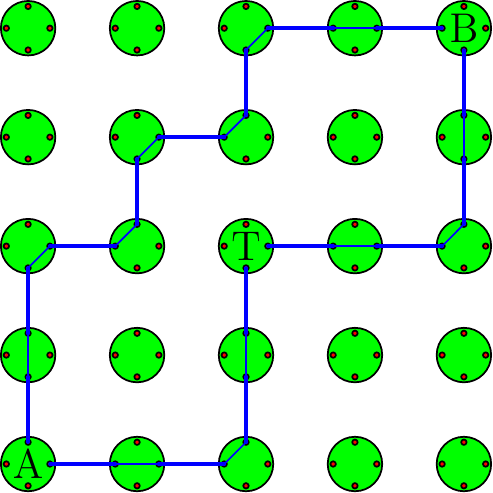}{Stage 2}{(a) The initial network of a $5\times5$ grid, with a central trusted node ($T$) between Alice ($A$) and Bob ($B$). (b) A possible network configuration after stage 1, in which some nodes succeed in establishing a shared state of the form in Equation (\ref{eq:shared-state0}) with their neighbors over the fiber channels and some do not. 
(c) Successful entanglement chaining and routing (depicted as the ``internal links'' at repeaters) as discussed in Stage 2. In this graph, routing has resulted in end-to-end paths being created between each pair of $A$, $B$ and $T$. These routes are successful only if every Bell state measurement along the path succeeds. Furthermore, the probability of decoherence increases with the path length in our model as discussed; thus minimizing the path lengths through effective routing algorithms and trusted nodes is advantageous.  See also Figure \ref{fig:guide}.}{stages1-3}


Users and trusted nodes perform the E91 QKD protocol.  For notation, we assume there are $n$ trusted nodes denoted $T_1, \ldots, T_n$.  We label Alice as $T_0$ and Bob as $T_{n+1}$.  For every pair of users and trusted nodes there exists a private raw-key pool; that is, for every $i,j \in \{0,\cdots, n+1\}$, with $i < j$, there exists a buffer $\rawkey_{i,j}$, storing the raw key shared between nodes $T_i$ and $T_j$.  Note that repeaters do not have key buffers.  As we are using the entanglement based E91 protocol, the first goal of the network is to establish joint entanglement between these pairs of nodes $T_i$ and $T_j$, ideally the Bell state $\ket{\Phi^+} = \frac{1}{\sqrt{2}}(\ket{00} + \ket{11})$.  This resource may be used to establish a secret key between users.

The quantum network we analyze operates in {\em rounds}, each round consisting of {\em three primary stages} (see examples in Figures \ref{fig:stages1-3} and \ref{fig:guide}).  The first stage, and part of the second stage, are standard in repeater-only quantum networks (see \cite{pant2019routing,chakraborty2019distributed}). As mentioned, nodes have at most four inputs into them (since the two users are at the corner of grid, they each only have two inputs); we assume that nodes have the capability of storing a single qubit in short-term memory for each of these inputs, through one complete round.  After the round is complete, any qubits not measured, are discarded.  That is, the memory capabilities of nodes are not sufficient to store a qubit through two or more rounds in our network.  Note that users in our network have the fewest demands on quantum memory.

In the first stage, adjacent nodes attempt to share an entangled pair, in particular the Bell state $\ket{\Phi^+}$, with one particle remaining at a local node and the other being transmitted to an adjacent node.  Due to fiber loss, this succeeds with a certain probability $p = 10^{-\alpha L/10}$ where $\alpha$ is the fiber attenuation coefficient (we use $\alpha = .15$ in our later simulations).  Furthermore, even if successful, the entangled pair may not be a Bell state $\ket{\Phi^+}$ but instead will, with probability $D$, depolarize and become completely mixed.  Ultimately, after this first stage, adjacent nodes $u$ and $v$ will either not have any qubit in their short-term memory (with probability $1-p$) or, with probability $p$, they will share a quantum state of the form:
\begin{equation}\label{eq:shared-state0}
\rho_{u,v} = (1-D)\kb{\Phi^+} + D \cdot \frac{I}{4},
\end{equation}
where $I$ is the identity operator on two qubits.  Note this $D$ may be used to model channel noise along with noise internal to the quantum memory of the repeater.  As is standard in repeater-network analyses \cite{pant2019routing,chakraborty2019distributed}, we assume that nodes are able to determine whether a qubit has arrived in their short-term quantum memory or if it is a vacuum.

In the second stage, a routing protocol, to be discussed (see Section~\ref{sec:routing}), is performed to decide how to 
effectively route entanglement between nodes $T_i$ and $T_j$ for $i \ne j$.  This routing protocol can take into account whether nodes have a qubit in their short-term memory.  The goal of this stage, and in particular the routing algorithm, is to determine a set of paths, the end-points of which are either users or trusted nodes, while the interior nodes of each path are quantum repeaters.  Given these paths, the quantum repeaters will perform entanglement swapping operations on the qubits that are locally held in their short-term memory, which, if successful, create ``virtual'' entangled links between the end-points of each path.
The entanglement swapping operations at the repeaters, which themselves consist of Bell measurements, succeed only with a certain probability that we denote as $B$.  We assume the network is able to determine whether a successful entanglement swapping occurred or not.  Thus, at the end of this stage, end-nodes $T_i$ and $T_j$ on each path can determine whether the network was successful in creating an end-to-end  path between these nodes.  Note that they cannot tell if the state they share is the correct Bell state $\ket{\Phi^+}$ or the completely mixed state $I/4$ due to depolarization.  At this point, repeaters are no longer needed until the next round, and users and trusted nodes share a state of the form in Equation (\ref{eq:shared-state}), where the probabilities of the mixed state depend on $D$ and the path length -- in particular, on a path consisting of $k$ interior nodes, the shared state will either be:
\begin{equation}\label{eq:shared-state}
\rho_{T_iT_j} = (1-D)^{k+1}\kb{\phi^+} + (1-[1-D]^{k+1})\cdot\frac{I}{4},
\end{equation}
or a vacuum if one or more repeaters failed in their entanglement swapping operation.

Finally, in the third stage, all pairs $T_i$ and $T_j$ of users and trusted nodes which have a shared state, and not a vacuum, attempt to distill a raw-key bit using the E91 QKD protocol.  If successful (which depends on users choosing the correct basis, and as such it is successful with probability $1/2$), the key-bit is added to a local raw key-pool for those users $\rawkey_{i,j}$.  Any qubit that was not measured by now is discarded and the network repeats at stage 1, attempting to establish fresh entangled pairs between adjacent nodes.  Refer to Figures \ref{fig:stages1-3} and \ref{fig:guide}.

We wrote a custom simulator to simulate the behavior of this network and these three stages.  This simulator maintains a raw-key pool $\rawkey_{i,j}$ between each pair of users $T_i$ and $T_j$ as would be done in an actual operation of this network.  In practice, nodes $T_i$ and $T_j$ will, once the length of the raw key in $\rawkey_{i,j}$ is sufficiently large, perform error correction and privacy amplification, leading to a secret key of size $|\secretkey_{i,j}|$.  In our simulations, we are interested in the theoretical performance of this network, and so we only do this process once at the end of the simulation and set $|\secretkey_{i,j}| = |\rawkey_{i,j}|\cdot (1-2h(Q))$ where $Q$ is the error in the raw key pool $\rawkey_{i,j}$ and where $h(x)=x\log(x)-(1-x)\log(1-x)$ is the binary entropy function. Again, as we are only interested in theoretical behavior, we compute $Q$ based on the actual simulator data; in practice $Q$ may be estimated by sampling from $\rawkey_{i,j}$.

Of course, the ultimate goal of the network is to maximize the secret key pool between Alice and Bob (i.e., $T_0$ and $T_{n+1}$).  If $n=0$ (i.e., there is no trusted node), at this point, we are done.  However for $n\ge 1$, a final key-routing process is required of the trusted nodes who must attempt to ``push'' a maximal amount of key material to Alice and Bob.  For $n=1$, this is trivial: $T_1$ will simply broadcast $\secretkey_{0,1}\oplus \secretkey_{1,2}$ (here, $T_0$ is Alice and $T_2$ is Bob); if they are not of equal length, then only the left-portion is broadcast.  Alice and Bob already have a secret key of size $|\secretkey_{0,2}|$ before this process.  Thus, finally, they end up with a secret key of total size $|\secretkey_{0,2}| + \min(|\secretkey_{0,1}|, |\secretkey_{1,2}|)$.  When $n > 1$ the situation is more complicated; in this work we use the max-flow algorithm to determine how many additional bits are appended to $\secretkey_{0,n+1}$.  Note that, at the end, not all key bits in trusted node pools may be usable if there is an insufficient amount of key material in ``matching'' nodes.  Later, when we evaluate performance of our routing protocols, we will look at the key-rate of the network, namely the size of the final secret key between Alice and Bob (after the trusted nodes perform this final key-routing process) divided by the total number of network rounds.

\triplefigbig{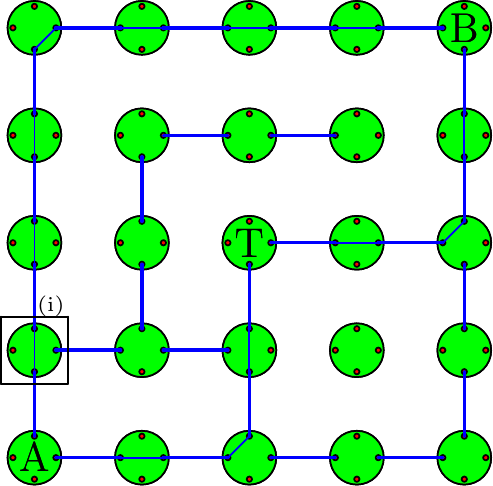}{Global algorithm.}{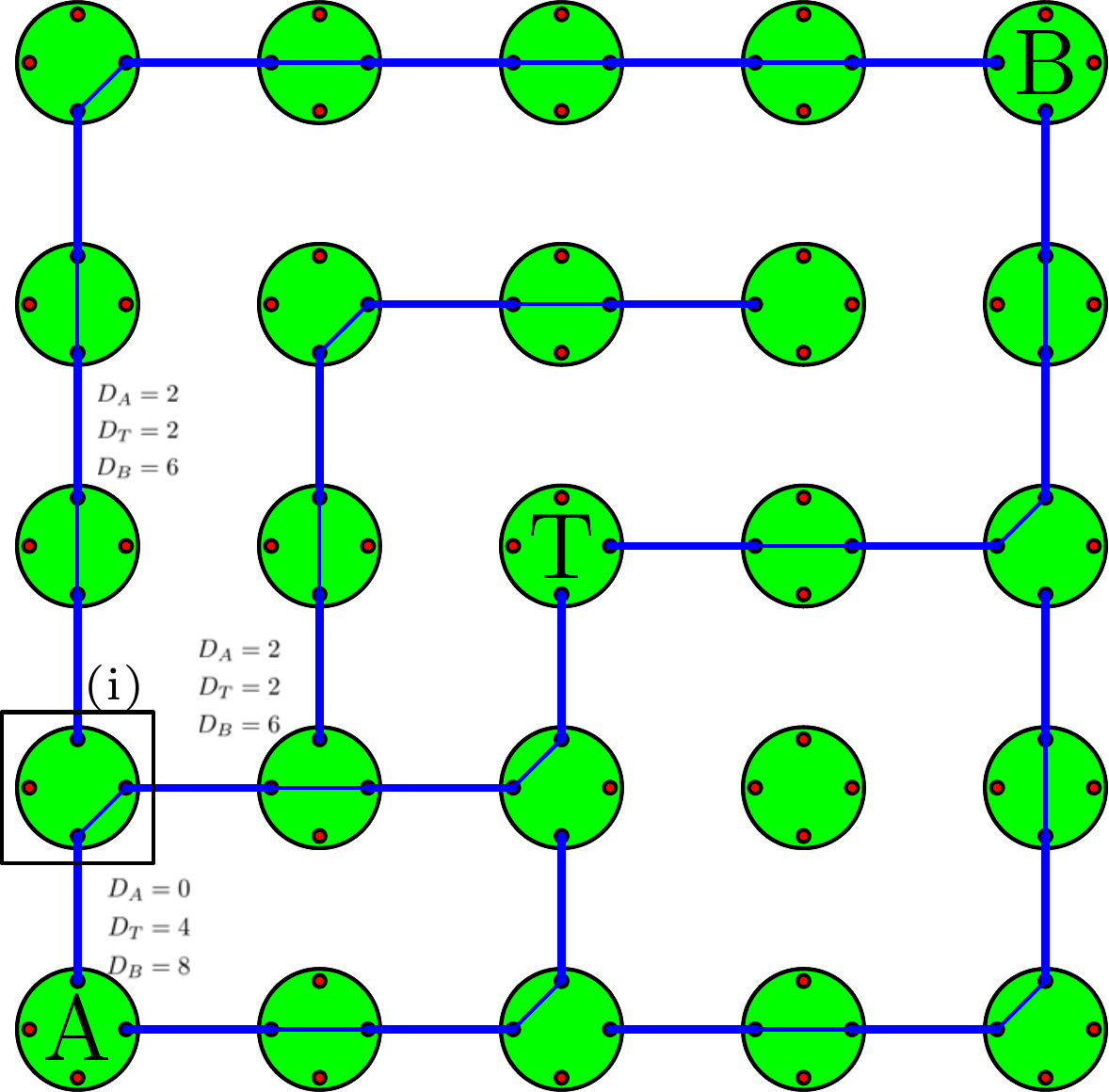}{Local algorithm: NIA}{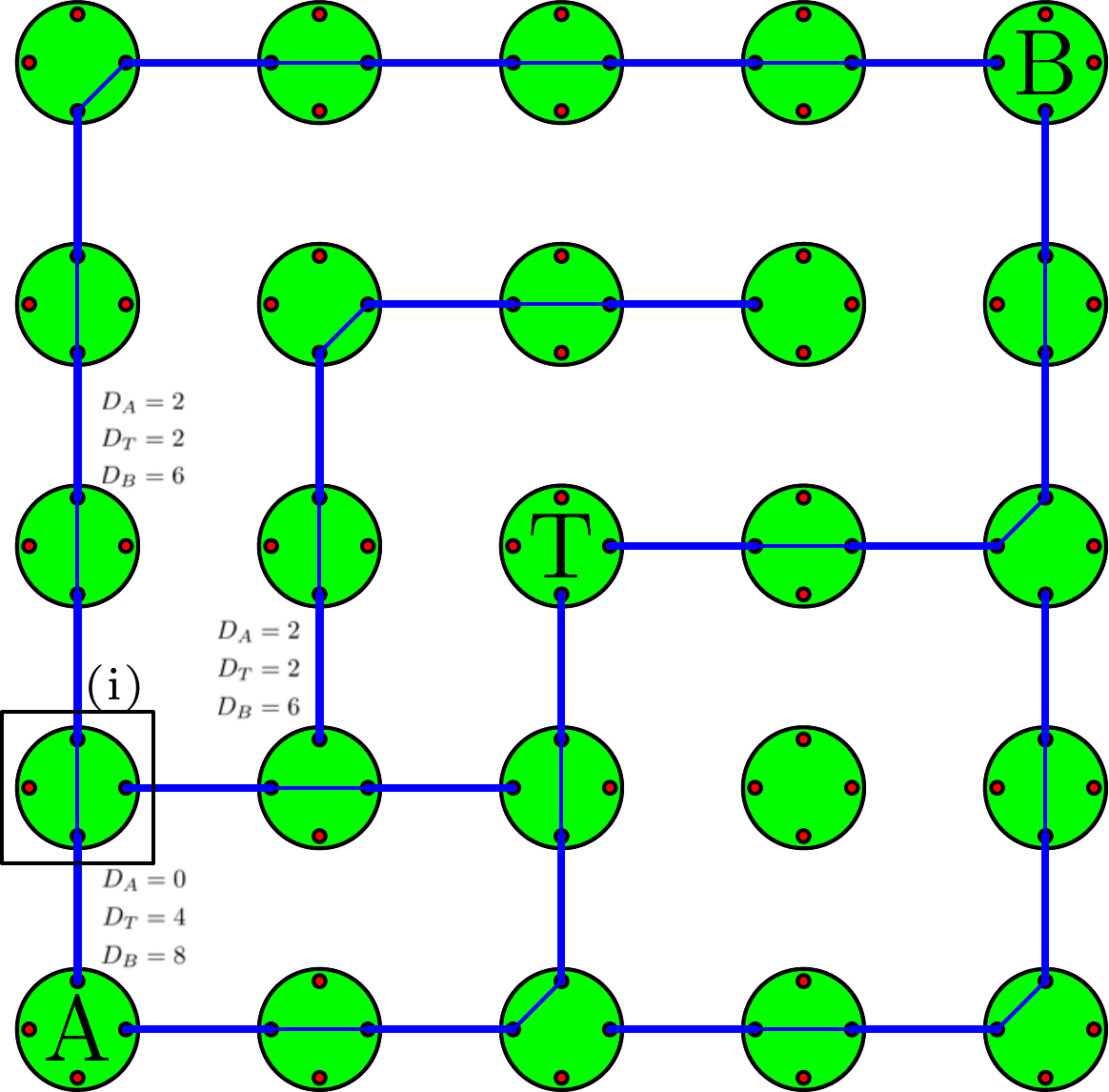}{Local algorithm: IA}{Illustration of the three routing algorithms, where $A$, $B$ and $T$ denote Alice,  Bob, and a trusted node respectively; the distances from a node to $A$, $B$ and $T$ are marked above the node. (a) The global routing algorithm finds three paths, between $A$ and $T$, $B$ and $T$, and $A$ and $B$, respectively. (b) The NIA algorithm finds two paths, one between $A$ and $T$, and the other between $B$ and $T$. The choice of the quantum repeater at location (i) cuts off a path between $A$ and $B$. (c) The IA algorithm finds three paths; the choice of the quantum repeater at location (i) allows forming a path between $A$ and $B$. }{Routing}

\section{Routing Protocols} \label{sec:routing}
In this section, we present three routing protocols that are used to form end-to-end paths for the second stage in a round.
These end-to-end paths have either users or trusted nodes as end nodes and have quantum repeaters as intermediary nodes.
We first present a high-level overview of these three protocols, and then describe each in detail.


\subsection{High-level Overview}
The three routing protocols are designed to (i) find short paths so that these paths can lead to higher chances of shared state (see Equation (\ref{eq:shared-state})), and (ii) find as many paths as possible to leverage the significant benefits of multipath routing over single-path routing~\cite{pirandola2016capacities,pirandola2019end} to lead to more key bits.

One of the three protocols relies on {\em global} link-state information, i.e., each node has the full knowledge of the external links in the network (i.e., the entanglement between pair of nodes that were created successfully in the first stage in a round). The other two protocols only require {\em local} link-state information, i.e., a node only knows 
which neighbors it shares a quantum state with and knows nothing beyond its neighbors. 
The network topology (including the locations of the users and the trusted nodes) is known for all protocols.  With the global knowledge, the global routing protocol is more likely to find better paths than the local routing protocols. On the other hand, gathering global information incurs longer latency since the link state information has to be propagated throughout the  network, while local link-state information is already available at the end of the first stage, and thus incurs no additional latency. As a result, the global routing protocol may only be applicable to small networks, while the local routing protocols can also be used in large networks.

Our routing protocols are inspired by work in \cite{pirandola2019end,pant2019routing}, where the authors consider routing between a pair of end users, without leveraging trusted nodes. We extend them to the scenarios with trusted nodes.
Of the two local routing protocols we develop, one significantly outperforms the other (see Section~\ref{sec:perf}), which can also be applied to scenarios with no trusted nodes in \cite{pant2019routing}.

\subsection{Global Routing Protocol}
The global routing protocol selects the shortest path (in hops) between any pair of nodes in $\{T_0, T_1,\ldots,T_{n+1}\}$, where $T_0$ is Alice, $T_{n+1}$ is Bob, and $T_1,\dots,T_n$ are trusted nodes. When two paths of equal length are found, one of them is selected randomly. Then all the links along the path are removed, and the procedure repeats for the remaining links until no path can be found. Fig.~\ref{fig:Routing}(a) illustrates this algorithm. In this example, there is only a single trusted node $T$ in the center of the network. The first shortest path found is the path (of 4 hops) between Bob and the trusted node $T$. After the links along this path are removed, the second shortest path (of 4 hops) found is the path between Alice and $T$. After that, a path (of 8 hops) between Alice and Bob is found.   


\subsection{Local Routing Protocols}
The two local routing protocols are closely related. In both algorithms, each repeater makes its own decisions on forming an ``internal link'' (i.e., performing  entanglement  swapping on two qubits in its short-term memory). We next describe the action at an arbitrary repeater $u$. For ease of exposition, we assume there is only a single trusted node $T$ in the network (the description below can be extended easily to multiple trusted nodes). Furthermore, without loss of generality, we assume $u$ has four neighbors, denoted as $u_a$, $u_b$, $u_l$ and $u_r$, corresponding to the nodes that are above, below, to the left of, and to the right of $u$, respectively. Let $D_A(u_a)$, $D_B(u_a)$ and $D_T(u_a)$ denote the distance (in hops) from $u_a$ to Alice, Bob and trusted node $T$, respectively. Similarly, define the distances for $u$'s other neighbors. 
Since each node knows the network topology, all the above distance values of $u$'s neighbors are known to $u$, which are used in $u$'s decision making. Specifically, in a given round, $u$ checks how many neighbors it shares quantum state with. If it shares quantum state with fewer than two neighbors, then it does nothing (i.e., no ``internal link'' can be created). If it shares quantum state with exactly two neighbors, it simply connects these two neighbors. Otherwise (i.e., it shares quantum state with more than two neighbors), it connects the two neighbors that have the shortest distances to two unique nodes including Alice, Bob and the trusted node. Afterwards, if there still remain two neighbors that $u$ shares quantum state with (i.e., $u$ established a quantum state with all four of its neighbors), then $u$ connects these two remaining neighbors.

The two local routing algorithms only differ in how they deal with ties. Specifically, if two potential sets of connections are equal in distance, the first local algorithm breaks the tie randomly, while the second algorithm favors horizontal or vertical ``internal links'', which has the effect of simplifying the paths that are created, and limiting the number of times one path utilizes a link that is integral to the formation of another path (see example below). We refer to the first local routing algorithm as \emph{Non-Intersection Avoidant (NIA)} algorithm, while we refer to the second one as \emph{Intersection Avoidant (IA)} algorithm. While these two algorithms only differ slightly, our simulation results in Section~\ref{sec:perf} demonstrate that the IA algorithm can significantly outperform the NIA algorithm. In fact, the performance of the IA algorithm approaches that of the global algorithm in some scenarios. 

Fig.~\ref{fig:Routing}(b) illustrates the NIA algorithm. In the example, the algorithm finds two paths, one between Alice and the trusted node $T$, and the other between Bob and $T$. At location (i), the repeater attempts to connect node $A$ (which has a minimum distance of 0 from Alice) with one of the other two neighbors, which have equal distance to the trusted node $T$. The algorithm selects one at random, which happens to be the neighbor to its right. While this ``internal link'' becomes part of a path from $A$ to $T$, it has a negative impact in that it cuts off any paths between Alice and Bob (or between Alice and $T$) that go through the nodes above (i). The above problem does not happen in the IA algorithm (see Fig.~\ref{fig:Routing}(c)), where the repeater at location (i) connects  the node above with node $A$, leading to a vertical ``internal link'', which becomes part of the longer path leading to Bob while another path is established to connect Alice with $T$. In summary, in this example the IA algorithm is able to establish a path between Alice and $T$, Bob and $T$, and a link between Alice and Bob, just as the global algorithm does, while the NIA algorithm lacks the logic to enforce this outcome. Our large-scale simulation in Section~\ref{sec:perf} demonstrates that the IA algorithm leads to statistically 
better performance than the NIA algorithm.

\section{Performance Evaluation} \label{sec:perf}

In this section, we evaluate the performance of quantum key distribution in quantum networks using the various routing algorithms. We will compare the performance of the algorithms in a number of settings and investigate the impact of various parameters on the performance.

\subsection{Evaluation Setup}
We consider the scenarios where Alice and Bob are placed at the two corners of an $N\times N$ grid.  
The parameters related to network topology include the size of the grid ($N$) and the length of the fiber channel of each edge (horizontal or vertical) of the grid ($L$). The quality of the quantum network is represented by the probability of a successful Bell state measurement (BSM) at a repeater ($B$), and the probability of decoherence ($D$, see Equation \ref{eq:shared-state}).   Note that $L$ is the fiber length between neighboring two nodes.  Since Alice and Bob are at opposite corners of the entire grid network, the actual distance, therefore, between the two users is actually $\sqrt{2}(N-1) L$.
Table~\ref{tab:defaults} summarizes these parameters, with their default values and range of the values.  


\begin{table}
    \centering
    \caption{Parameters explored in performance evaluation. }
        \begin{tabular}{| c | c | c | p{3.8cm}|} 
        \hline
        \textbf{Parameter} &  \textbf{Default} & \textbf{Range} & \textbf{Description} \\ \hline
        $N$ & 5    & 5 - 15       & The network consists of $N^2$ nodes arranged in an $N \times N$ grid. \\ \hline
        $L$ & 1 km & 1 - 20 km & The length of the fiber connecting neighboring nodes. The distance between Alice and Bob scales with both $N$ and $L$.  \\ \hline
        $B$ & .85  & 0.65 - 1.00  & Represents the quality of the repeaters. A Bell state measurement succeeds at a repeater with probability $B$. \\ \hline
        $D$ & .02  & 0.00 - 0.06  & Represents the amount of noise in the channels. EPR pairs decohere with probability $D$. \\
        \hline
        \end{tabular}
    \label{tab:defaults}
\end{table}

The performance metric is {\em key rate}, i.e., the average number of secret key-bits generated between Alice and Bob per round of network use as discussed in Section \ref{sec:model}. Unless otherwise stated, the number of rounds we simulate for each test and for each setting is $10^6$. We also evaluate the network's performance when there are no trusted nodes; a single trusted node; and two trusted nodes, along with different location configurations.

\subsection{No Trusted Nodes}

\begin{figure}[t]
	\centering
	\subfigure[Impact of fiber length.]{
		\includegraphics[width=0.24\textwidth]{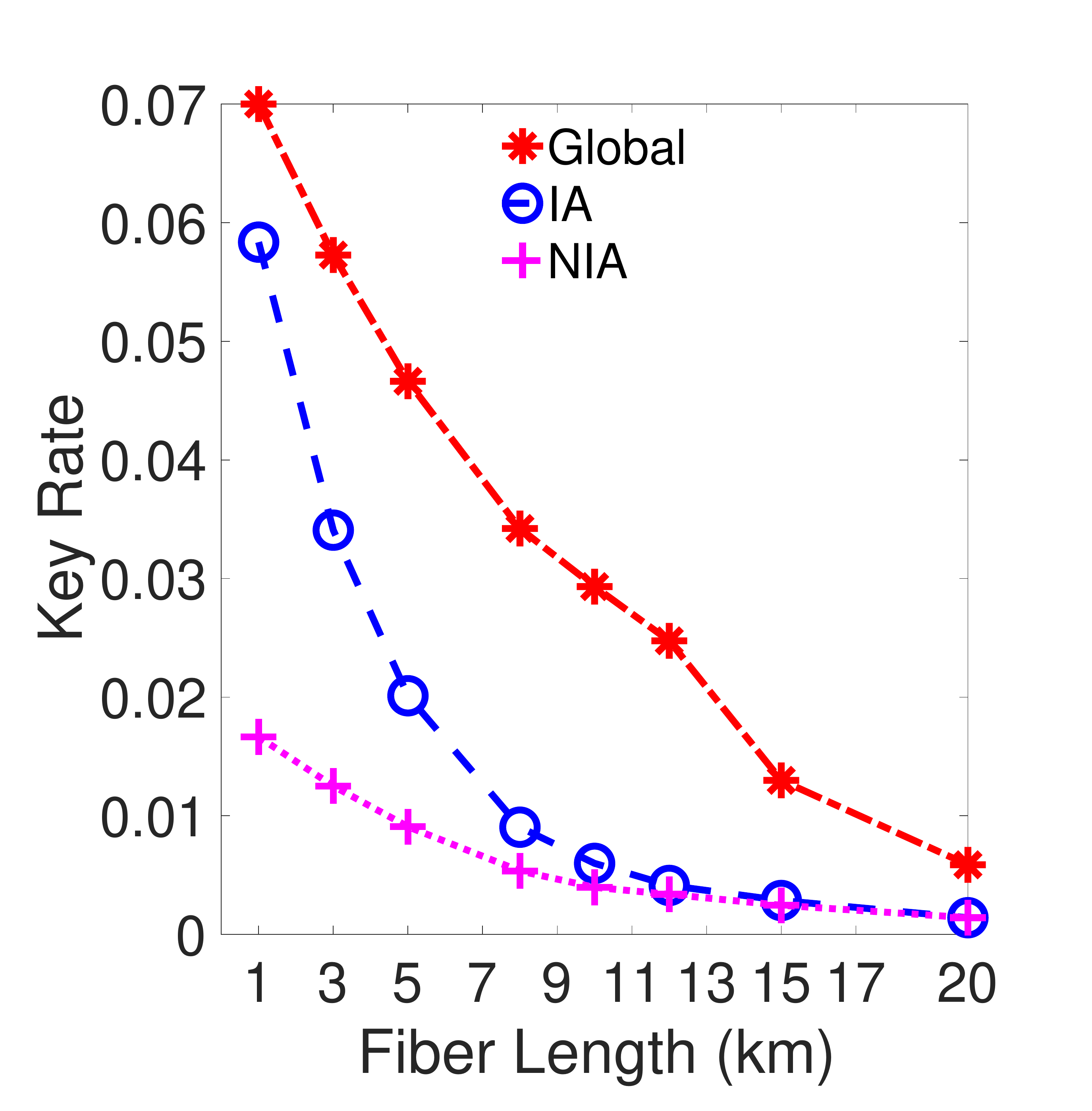}
	}
	\hspace{-.148in}
	\subfigure[Impact of decoherence rate.]{%
		\includegraphics[width=0.24\textwidth]{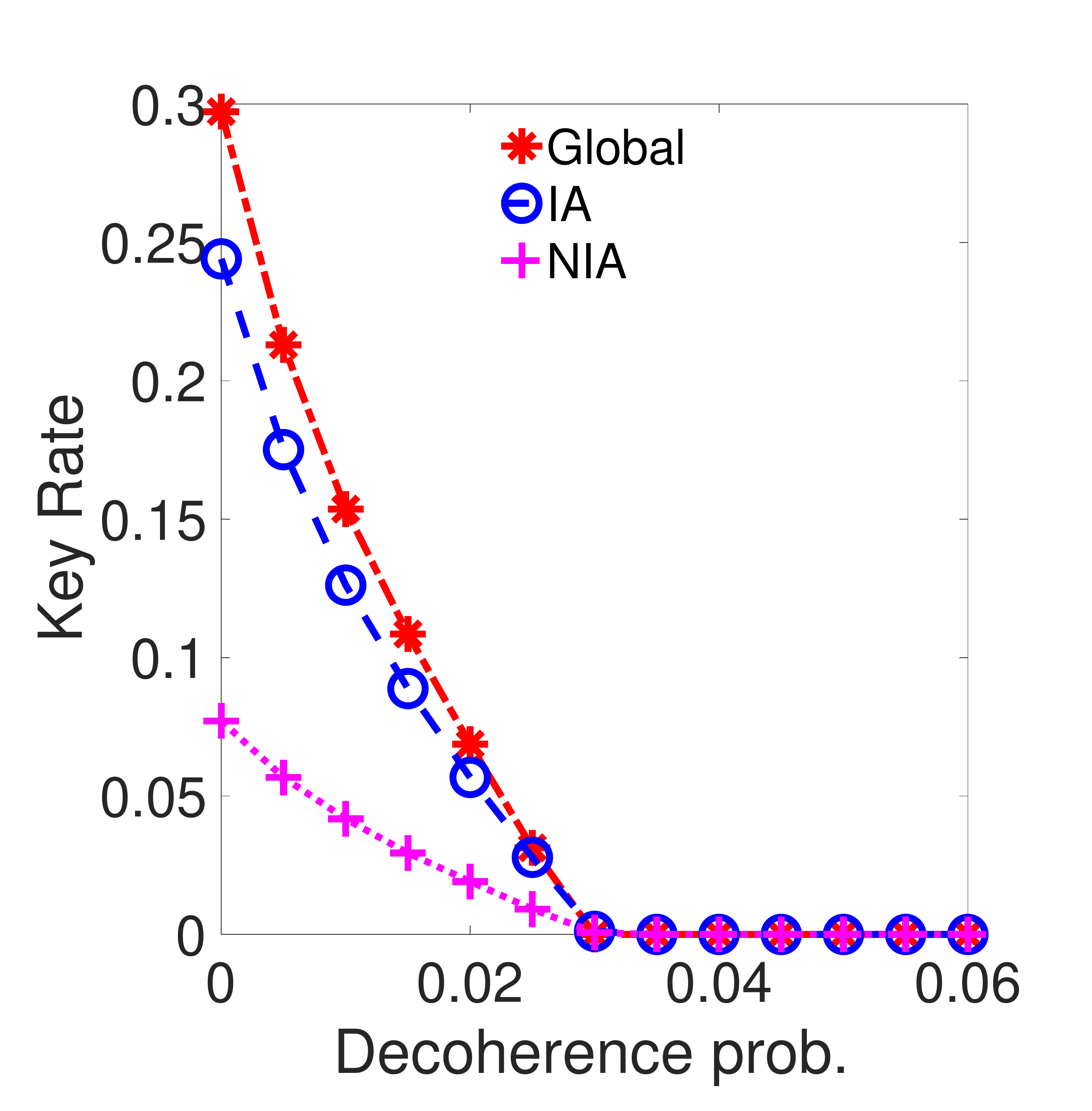}
	} \\
	\subfigure[Impact of BSM success rate.]{%
		\includegraphics[width=0.24\textwidth]{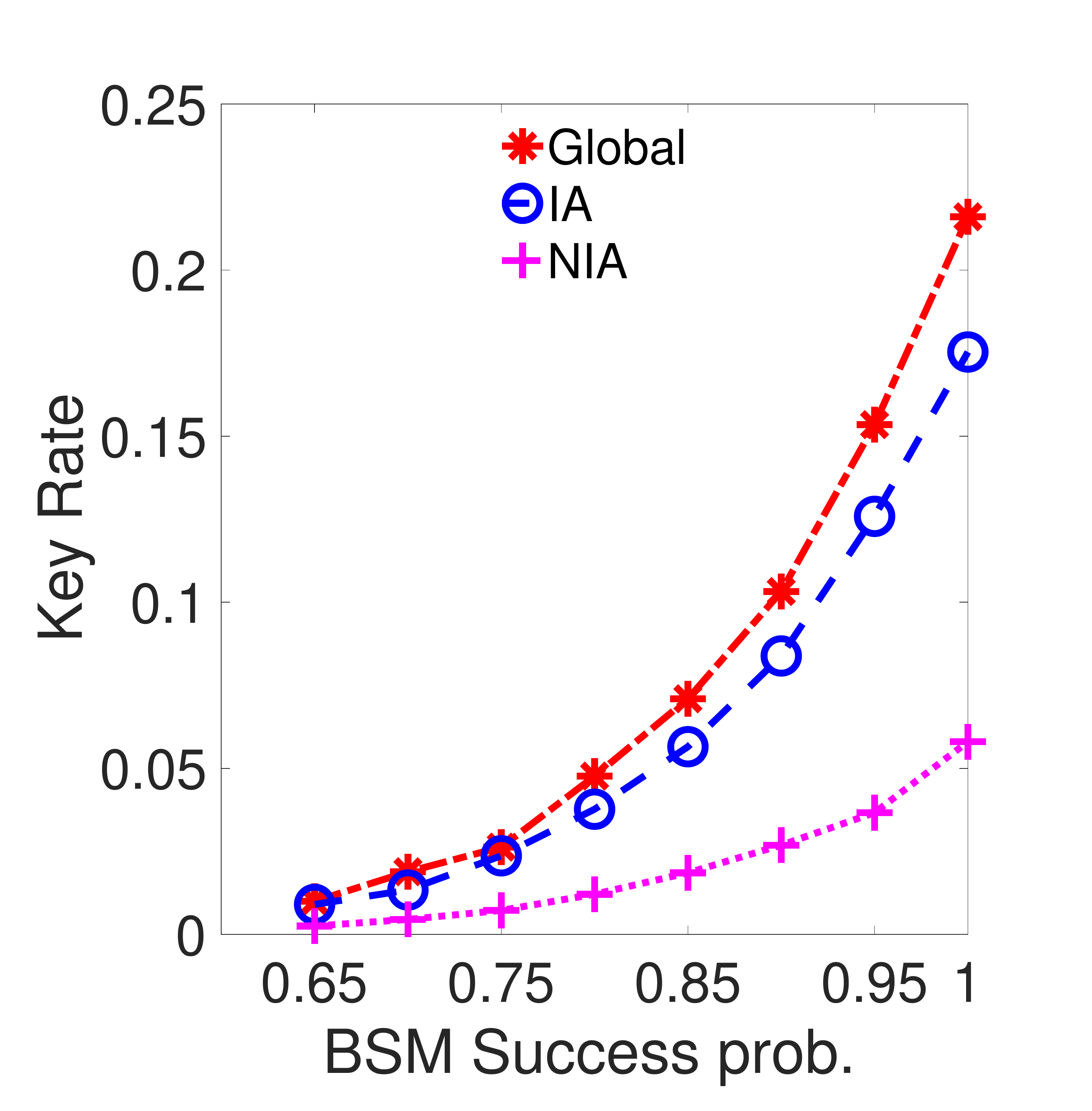}
	}
	\hspace{-.148in}
	\subfigure[Impact of network size.]{%
		\includegraphics[width=0.24\textwidth]{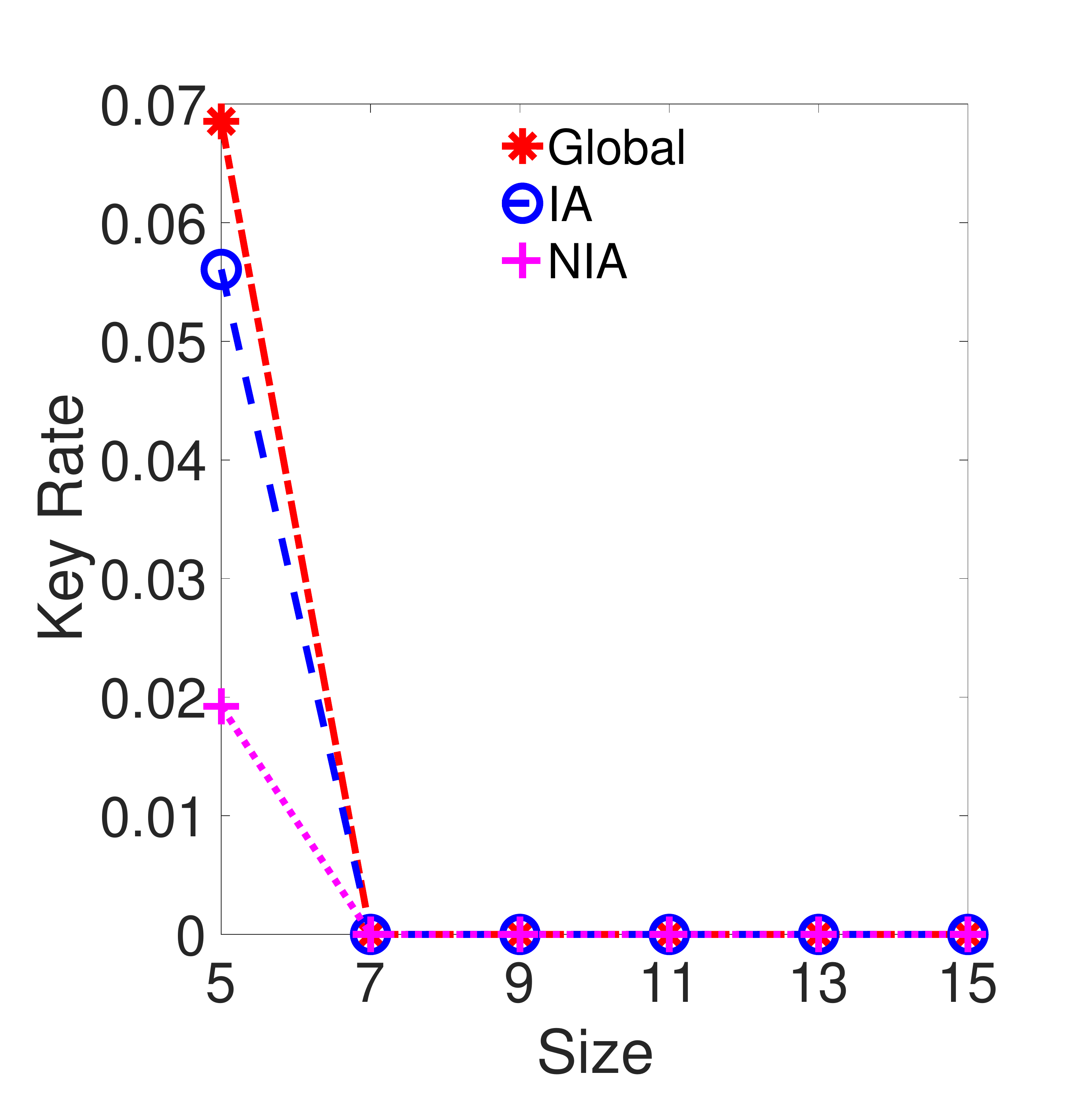}
	}
	\caption{Impact of various parameters on key rate when there is no trusted node in the network ($5 \times 5$ grid). Note, as discussed, ``fiber length'' $L$ is the length between nodes.  Thus, the actual distance between the two users is actually $\sqrt{2}(N-1) L$.}
	\label{fig:NoTNGraphs}
\end{figure}

We first investigate the scenario with no trusted node and only repeaters. These results serve as baselines to demonstrate the benefits of having trusted nodes in the network in Sections~\ref{sec:one-trusted-node} and \ref{sec:more-trusted-nodes}. Fig.~\ref{fig:NoTNGraphs}(a) plots the key rate when varying the fiber length, $L$, for a $5 \times 5$ grid. The results of the three routing algorithms are plotted in the figure. As expected, since increasing the fiber length leads to higher fiber loss, the key rate decreases with $L$. The global routing algorithm, also as expected, outperforms the two local algorithms. Between the two local algorithms, the intersection-avoidance (IA) algorithm significantly outperforms non-intersection-avoidance (NIA) algorithm, particularly for lower fiber lengths.
The difference between the global algorithm and IA algorithm is small for short fiber length, and increases when the fiber length increases, where having global knowledge allows the algorithm to find paths, even when they are longer and more complicated.

Fig.~\ref{fig:NoTNGraphs}(b) plots key rate when increasing the decoherence rate $D$. We again observe that the global routing algorithm outperforms the two local algorithms, and the IA algorithm significantly outperforms the NIA algorithm.
As expected, increasing the decoherence rate leads to lower key rates; the decrease is particularly dramatic when increasing the decoherence rate from 0 to 2\%. As an example, for the global routing algorithm, the key rate decreases from .3 key-bits/round to just over .07 key-bits/round when increasing the decoherence rate from 0 to 2\%.
We also note that, interestingly, while both the global and IA algorithms achieve a key rate around .07 at 2\% decoherence, the NIA algorithms, under the ideal condition of 0\% decoherence, achieves the same key rate of .07 key-bits/round. 
The drastic advantage of the IA algorithm over the NIA algorithm highlights the importance of designing effective local routing algorithms.

Fig.~\ref{fig:NoTNGraphs}(c) plots key rate when varying $B$, the probability that the Bell state measurement at a repeater succeeds. The relative performance across the three routing algorithms is similar as the above two scenarios. 
Of note, it is evident 
that increasing the reliability of these measurements can lead to a significantly improved key rate. At a 75\% success rate, for example, the global algorithm achieves just under .026 key-bits/second, but an increase to 95\% or even 100\% success rates allows us to achieve key rates of approximately .153 and .216 respectively. 

Last, Fig.~\ref{fig:NoTNGraphs}(d) demonstrates the impact of network size $N$ on key rate.  Note that, by increasing $N$ (the number of nodes in the network) but keeping the horizontal/vertical distance between two adjacent nodes fixed at 1 km, we are effectively increasing the total distance between Alice and Bob (they are at the two corners of the grid). For all the routing algorithms, the key rate decreases when the network size (i.e., total distance) increases. This is expected since, as the network size increases, the path between Alice and Bob becomes longer,
and therefore the overall error (see Equation \ref{eq:shared-state}) and probability of a path failing increases, leading to lower key rate. Of the three routing algorithms, the global algorithm is less affected compared with the other two algorithms since it is more capable of finding longer paths as the network size increases than the two local algorithms.

We observe that, as the distance between the end users $A$ and $B$ increases, even using quantum repeaters with perfect internal performance (i.e., $B = 1$), the efficiency of the entire network degrades rapidly (we shall discuss this further in in Fig.~\ref{fig:NoiseLevels}).  We show in the following sections how even a single trusted node can dramatically improve the performance of QKD, indicating that, even in the future as repeater technology becomes more prevalent, trusted nodes may still be necessary to ensure high key rate when the distance between end users is significant.

\begin{figure}[t]
	\centering
	\subfigure[Impact of fiber length.]{
		\includegraphics[width=0.24\textwidth]{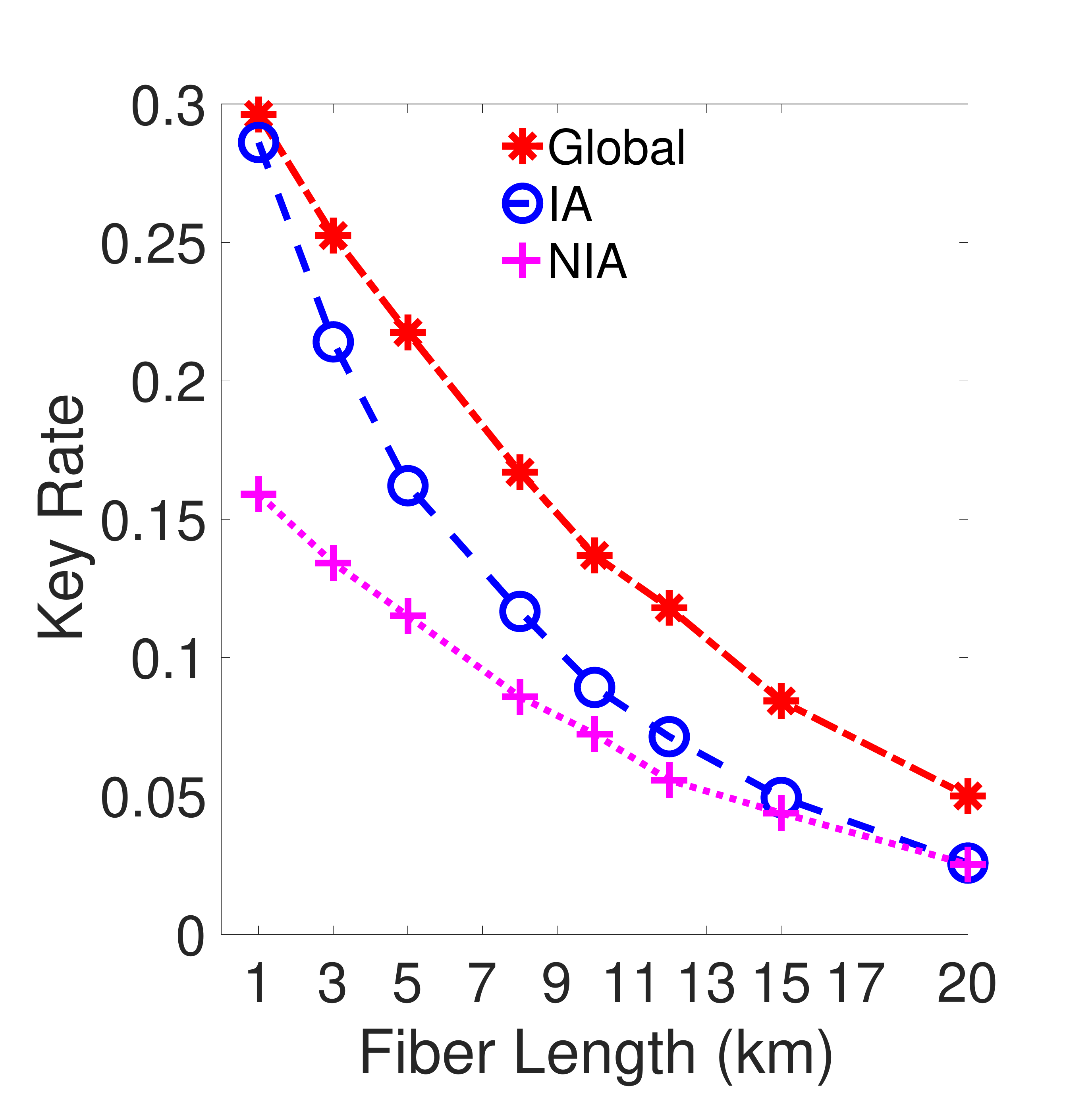}
	}
	\hspace{-.148in}
	\subfigure[Impact of decoherence rate.]{%
		\includegraphics[width=0.24\textwidth]{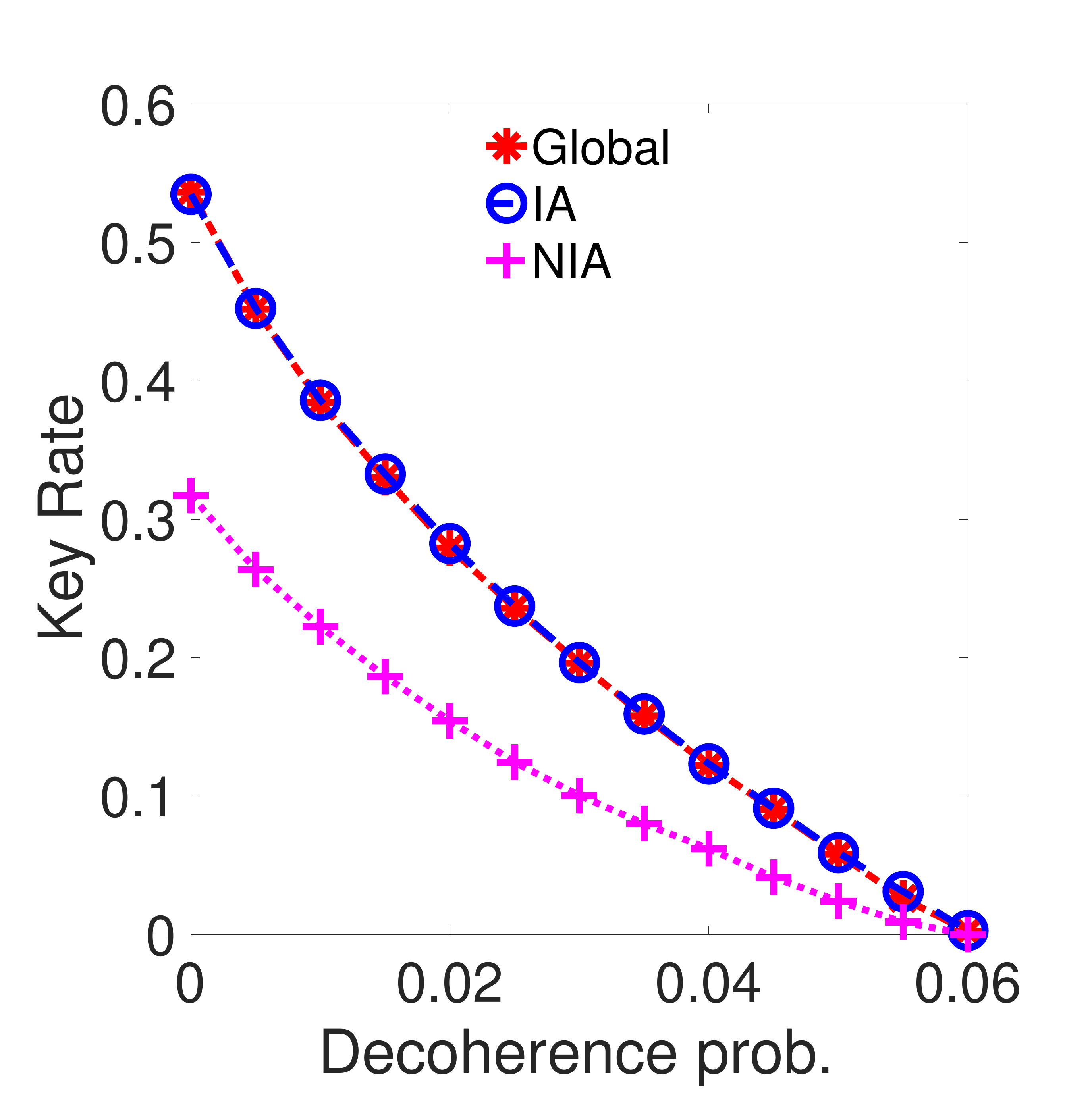}
	} \\
	\subfigure[Impact of BSM success rate.]{%
		\includegraphics[width=0.24\textwidth]{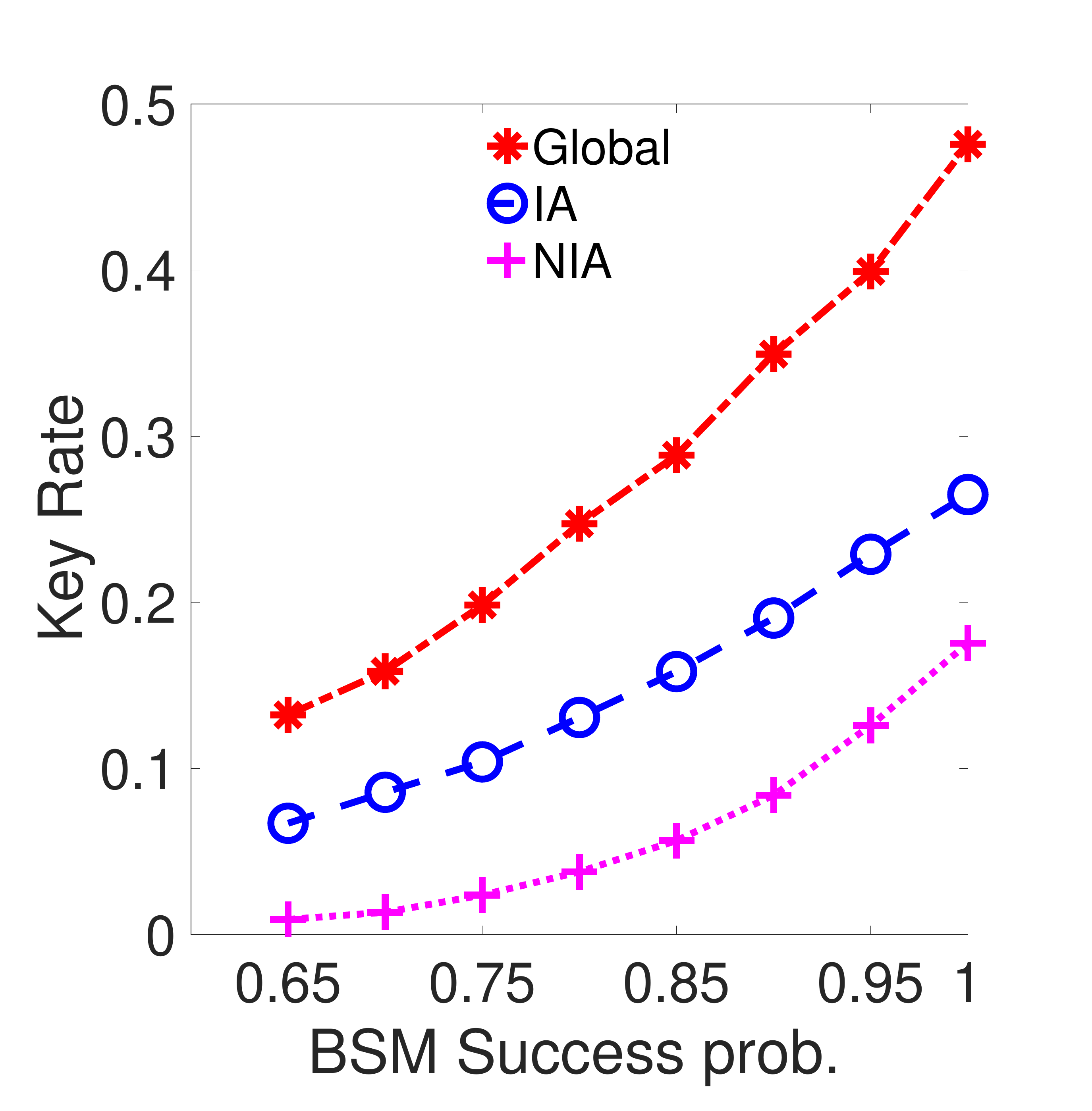}
	}
	\hspace{-.148in}
	\subfigure[Impact of network size.]{%
		\includegraphics[width=0.24\textwidth]{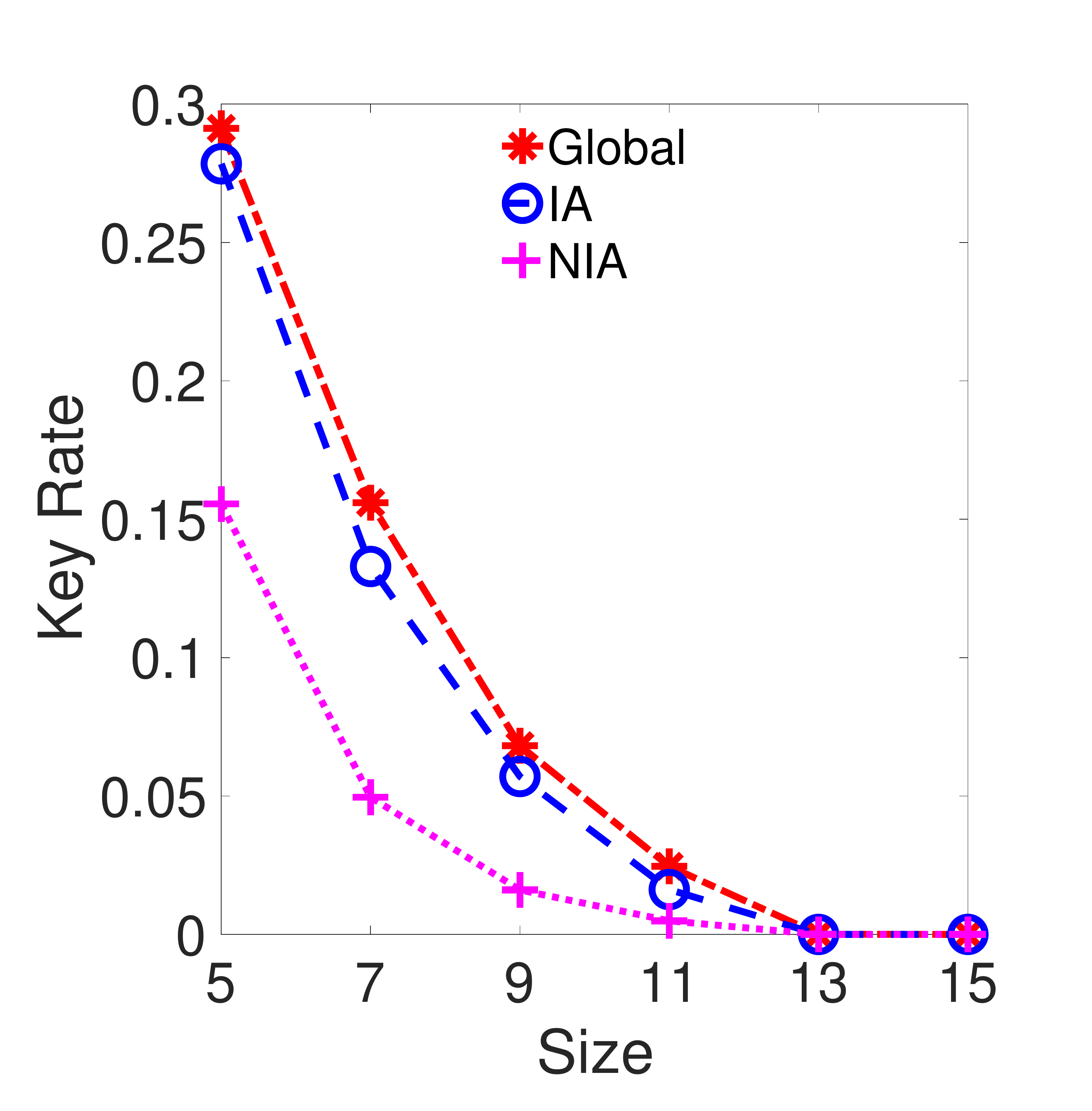}
	}
	\caption{Impact of various parameters on key rate when there is one trusted node in the network ($5 \times 5$ grid).}
	\label{fig:1TNGraphs}
\end{figure}

\subsection{Single Trusted Node} \label{sec:one-trusted-node}
We now consider the scenario where a single trusted node is placed at the center of the network grid, and investigate the impact of various parameters on key rate in this scenario. Figures~\ref{fig:1TNGraphs}(a)-(d) plot the key rate when varying the fiber length, decoherence rate, BSM success rate and network size, respectively. The results show that, under all three routing algorithms, having one trusted node leads to significantly higher key rate than the same setting with no trusted node (see Fig.~\ref{fig:NoTNGraphs}).
As an example, when there is no trusted node, as shown in Fig.~\ref{fig:NoTNGraphs}(a), the maximal key rate achieved for the global algorithm was .07 key-bits/round, while it is .3 key-bits/round, more than $3\times$ higher, as shown in Fig.~\ref{fig:1TNGraphs}(a) when there is one trusted node. 

Comparing Fig.~\ref{fig:1TNGraphs}(a) and Fig.~\ref{fig:NoTNGraphs}(a), we see that for the global algorithm,  when varying the fiber length from 1 to 15 km, the key rates with one trusted node are $.08$-$.25$
higher than that with no trusted node under the same setting, corresponding to a $1.9$-$4.4\times$ increase; the corresponding increases in key rate for IA and NIA algorithms are $1.2$-$8.8\times$ and $2.1$ - $9.2\times$, respectively. 
In fact, with one trusted node and in the same setting, the NIA algorithm achieves greater key rates than even the global algorithm could with no trusted nodes. This is because the trusted node essentially reduces the ``size'' of the network, allowing shorter paths to be constructed between Alice/Bob and the trusted node, As the local algorithms have information on a higher fraction of the total network for smaller networks, this reduction in the effective size of the network has a particularly strong effect on the local algorithms. 
This trend is also exhibited in Figs.~\ref{fig:1TNGraphs}(b)-(d). Where, again comparing with their counterparts in Fig.~\ref{fig:NoTNGraphs}, we see that the addition of a trusted node results in an almost two-fold increase in maximum achievable key-rate for the cases of decoherence, BSM success rate and network size.

In Fig.~\ref{fig:NoiseLevels}, we show the performance of the three algorithms with no trusted nodes and one trusted node in two additional channel scenarios: perfect BSM success probability with 2\% decoherence and 85\% BSM success probability with 5\% decoherence. In  Fig.~\ref{fig:NoiseLevels}(a), we see that even with the elimination of BSM failures, as the length increases between nodes the key rate still drops significantly. We see that even the NIA algorithm with a central trusted node (TN) outperforms the global algorithm with no TN, as the shorter paths between the parties and the TN are simply more likely to exist than the longer paths that must exist to connect Alice and Bob in the no TN case. In  Fig.~\ref{fig:NoiseLevels}(b) we consider a less idealized scenario, in which the BSM probability is our default of $85\%$ and our decoherence rate is higher, at $5\%$. We see that for this scenario, not even the global algorithm can achieve a positive key rate for the no TN case (at least in $10^6$ rounds of the network), and that in general the key rates achieved by the central TN is significantly lower (between $5-10\times$) than was achieved with $2\%$ decoherence in Fig.~\ref{fig:1TNGraphs}(a). It is notable, however, that in Fig.~\ref{fig:NoiseLevels}(b) we see that using 1 TN, each algorithm is able to maintain a positive key rate even when there is a 5\% decoherence rate, and Alice and Bob  are separated by 113 km (when $L=20$ km, considering a $5\times5$ network grid).

\tydubfigsingle{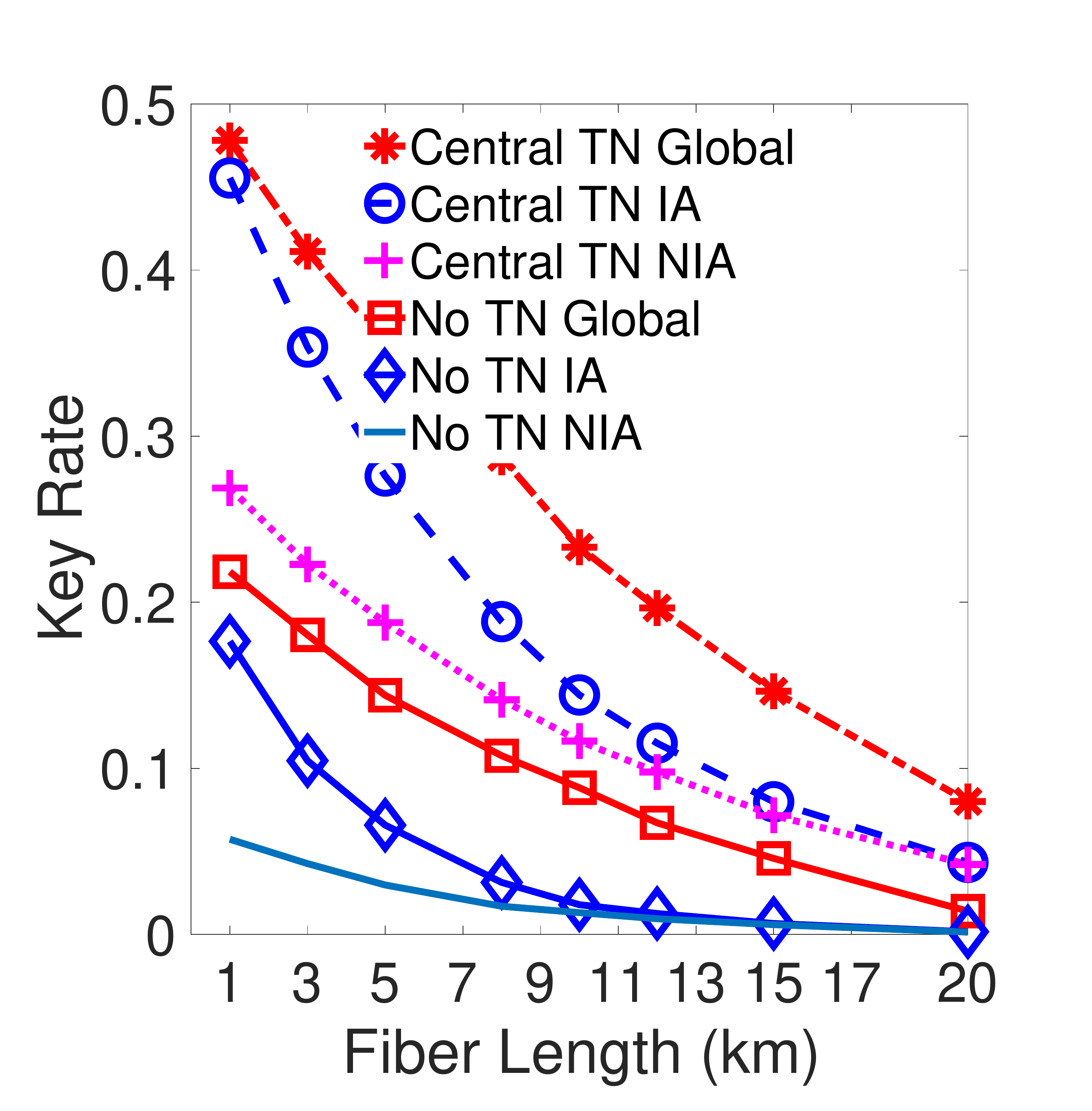}{100\% BSM Success Prob.}{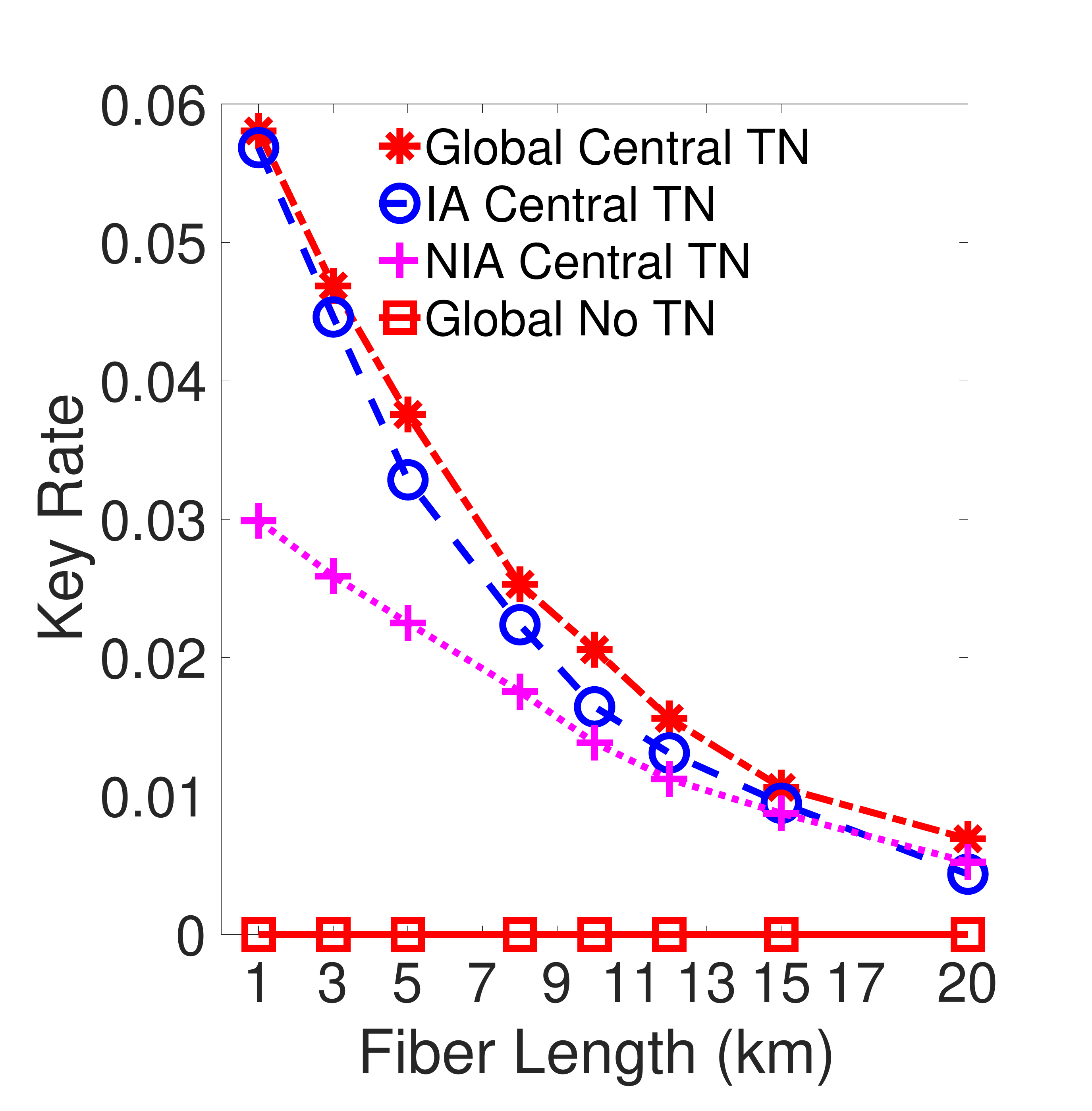}{5\% decoherence rate}{(a) The capabilities of the network when operating in an idealized scenario with no BSM failures (i.e. perfect repeaters) and 2\% decoherence probability. (b) The capabilities of the network with 5\% decoherence probability and 85\% BSM success probability.}{NoiseLevels}

\subsection{Multiple Trusted Nodes} \label{sec:more-trusted-nodes}


\singlefig{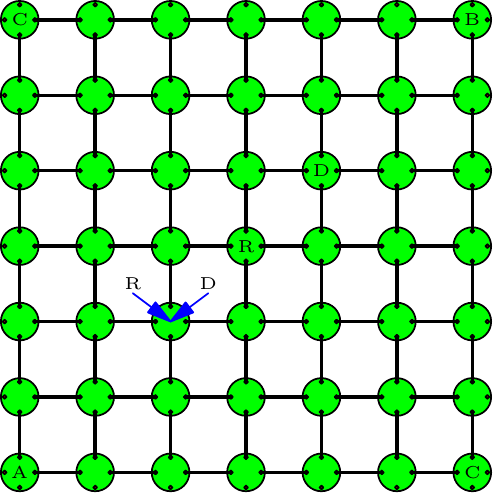}{A $7\times7$ network with 3 different ways of placing the two trusted nodes. The nodes labeled with $A$ and $B$ are Alice and Bob in each type of placement; the two nodes labeled with $C$ are the trusted nodes in the {\em corner} placement; the two nodes labeled with $D$ are the trusted nodes in the {\em diagonal} placement; and  the two nodes labeled with $R$ are the two trusted nodes in the {\em asymmetric} placement.}{TNPlacement}


\triplefigbig{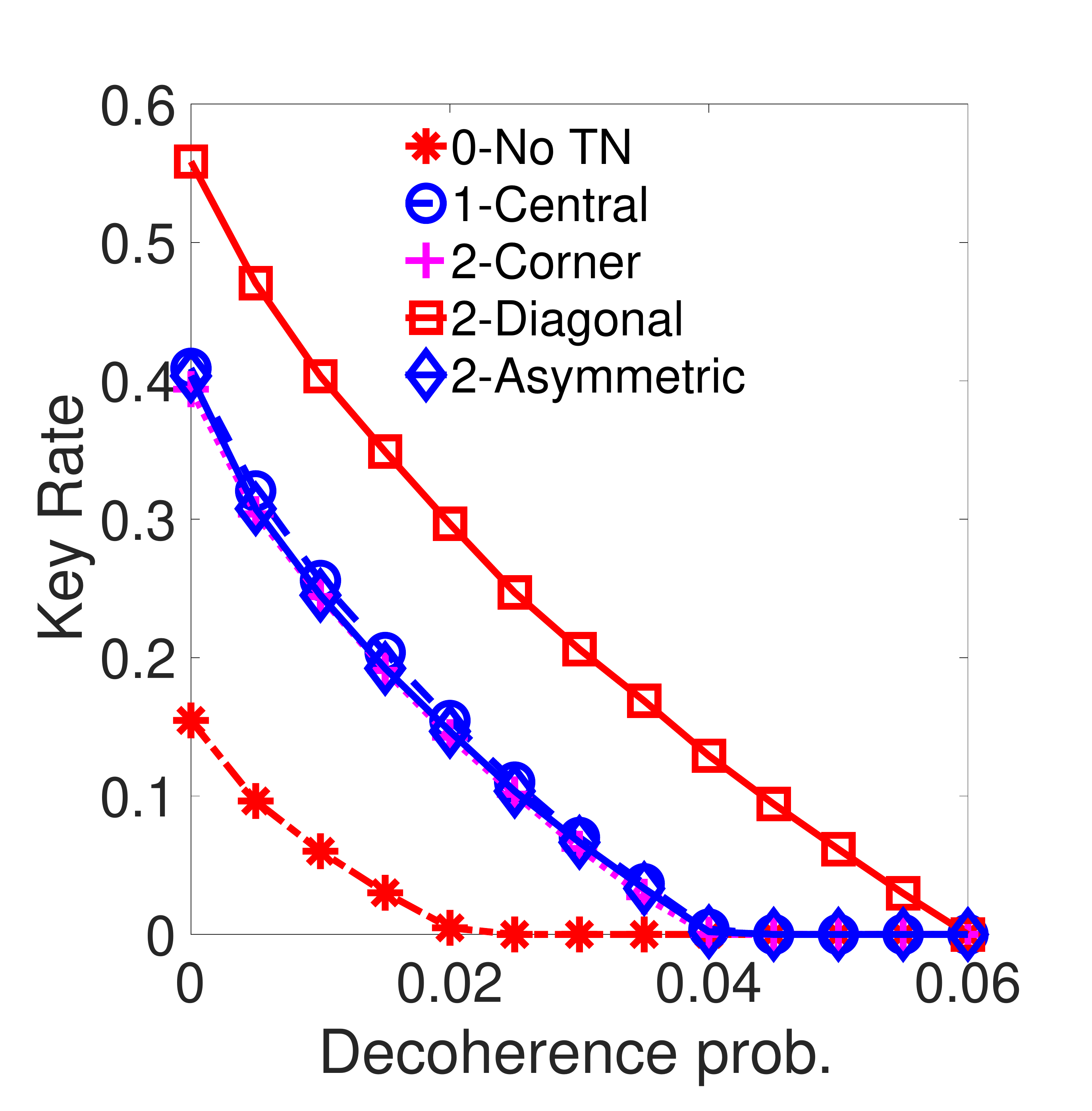}{Global}{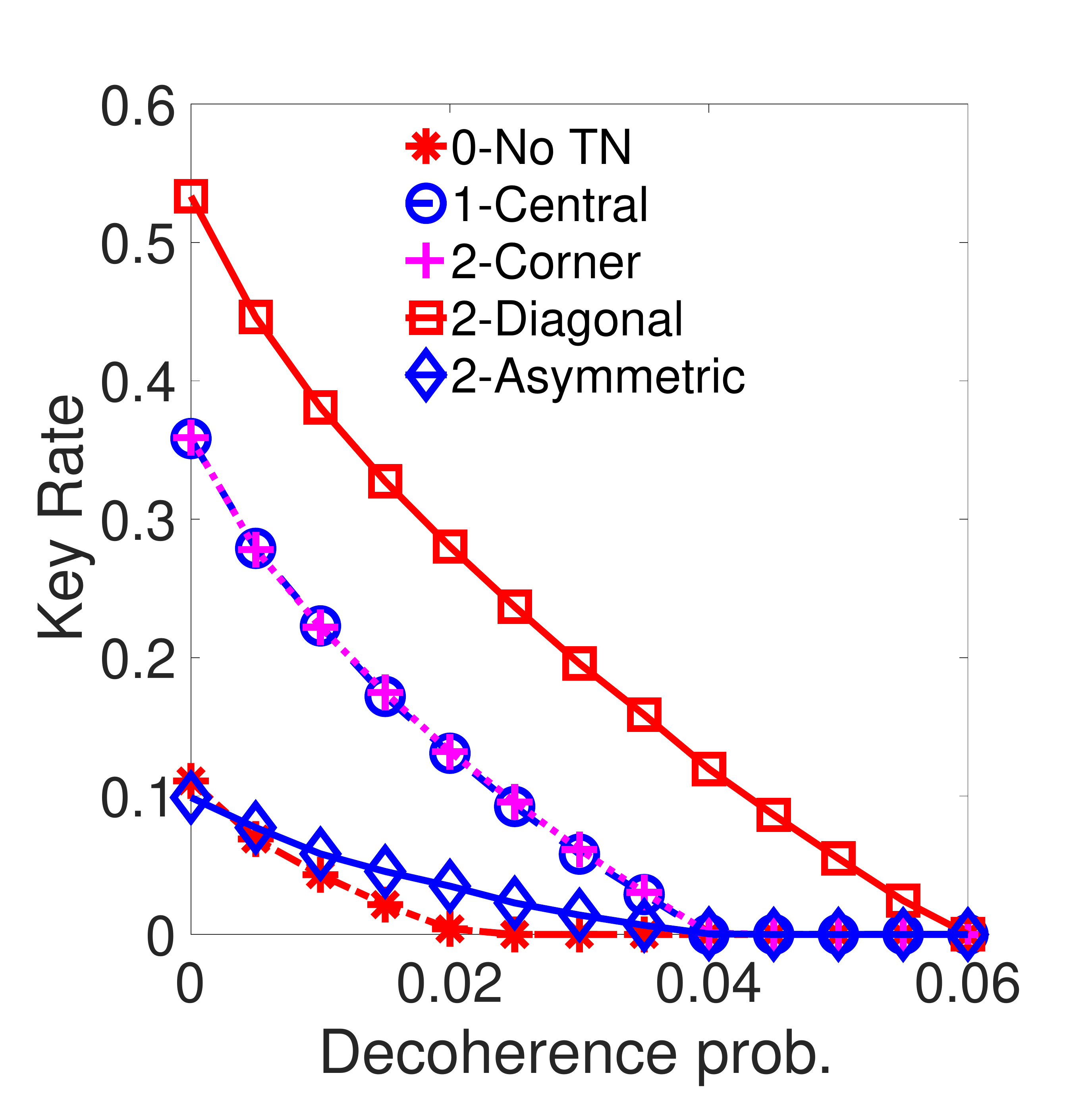}{Intersection Avoidant}{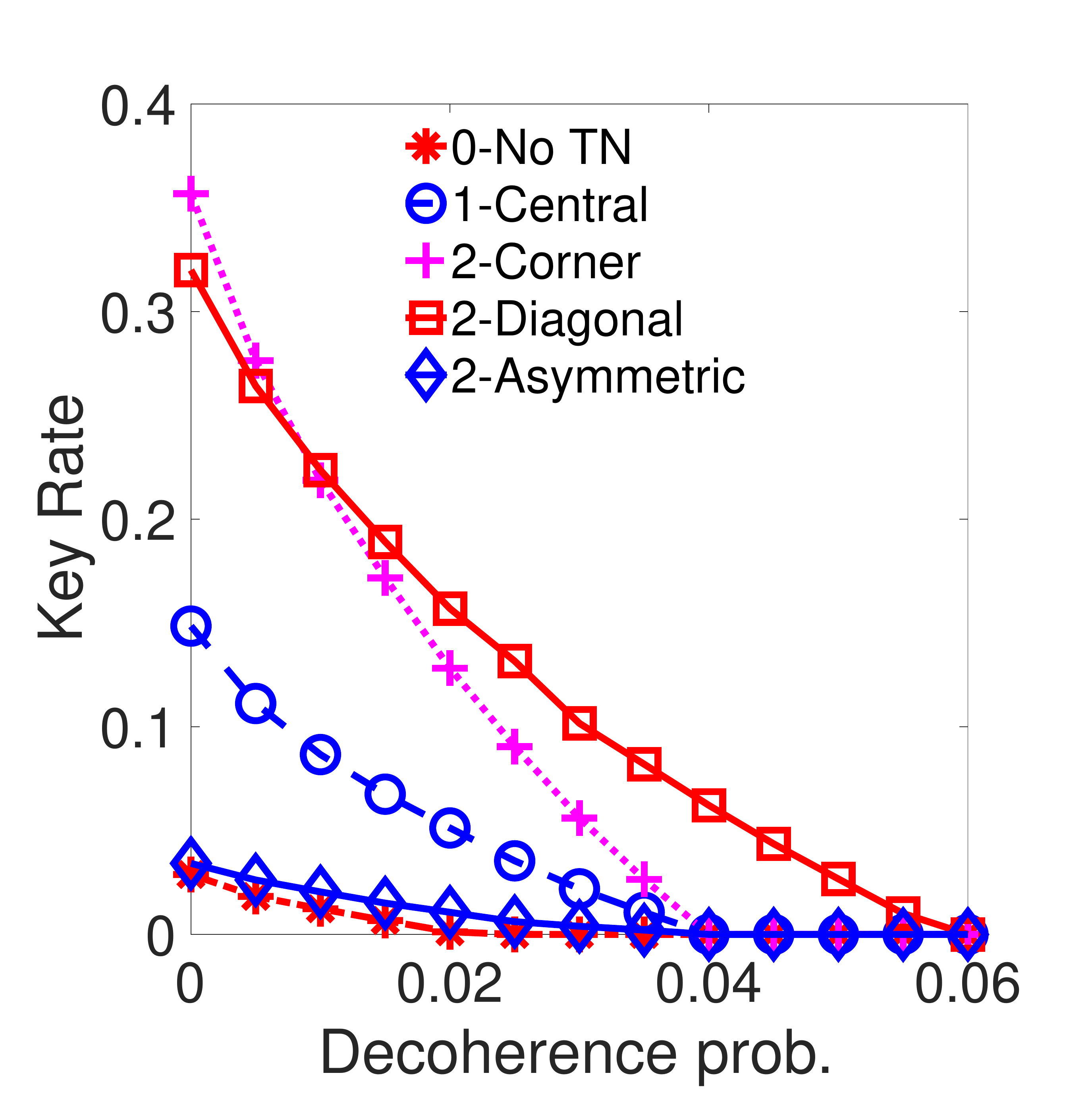}{Non-Intersection Avoidant}{Key rate versus the decoherence rate for the three routing algorithms in a $7\times7$ grid with $L=1$ km and $B=.85$. We consider the No TN case, the Central TN case, and the 2 TN cases in which there are two trusted nodes in the network, following corner, diagonal or asymmetric placements. }{2TNGraphs}

\tydubfigsingle{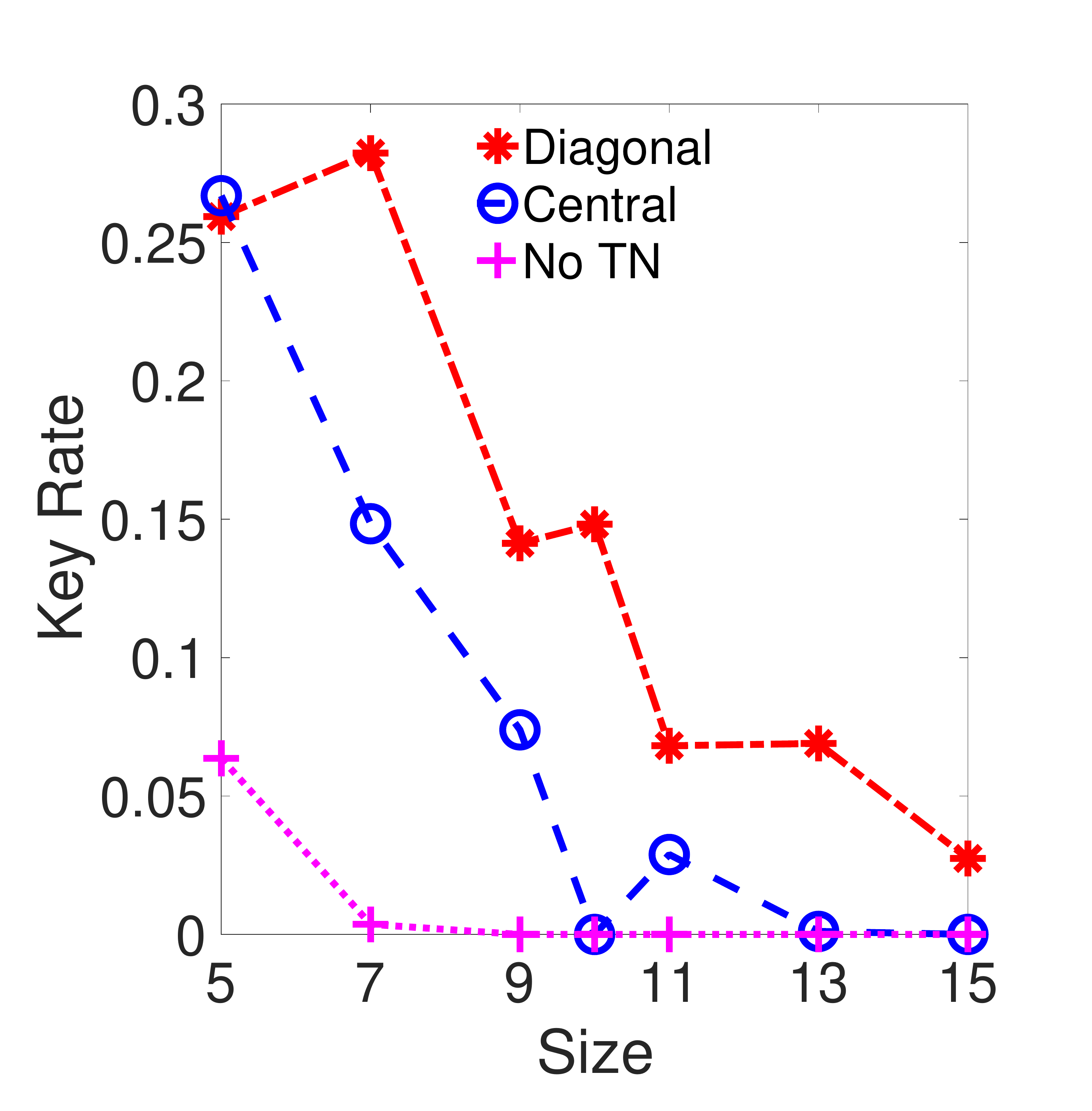}{Global}{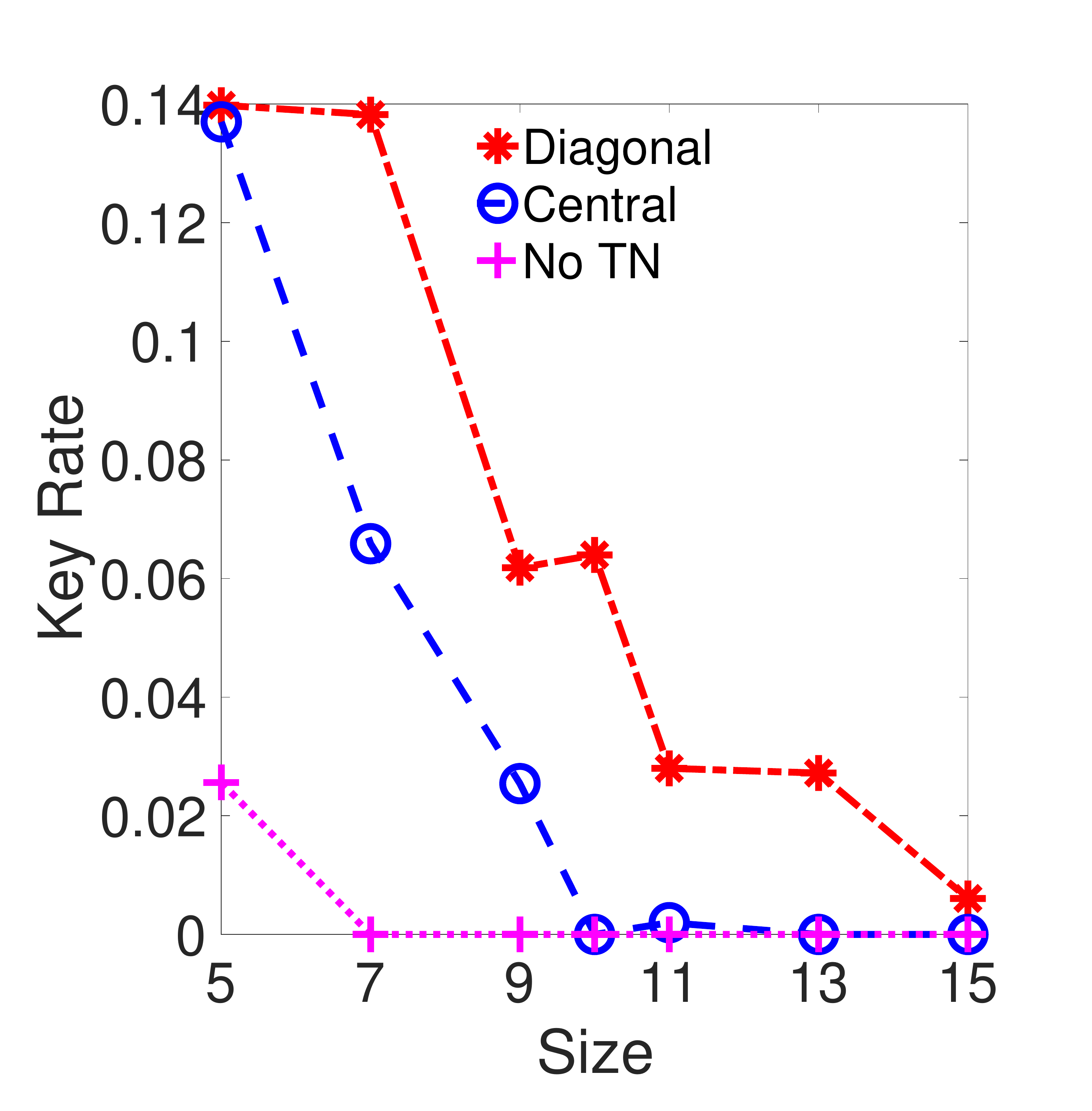}{Intersection Avoidant}{Key rate versus the network size for the global and IA routing algorithms for a 10 km $\times$ 10 km network grid where the fiber length between each node decreases as the network size increases. Notably, we see some non-monotonicity as the diagonal configuration does not operate effectively at grid sizes that are not of the form $3x+1$ for some integer $x$, and the central case does not operate effectively at even grid sizes such as 10x10. }{FixedLength}

\tydubfigsingle{figs/B_all2E_global}{Global}{figs/B_all2E_IA}{Intersection Avoidant}{Key rate versus the BSM Success rate for the global and IA routing algorithms in a $7\times7$ grid with $L=1$ km and $D=.02$. We consider the no TN case, the central TN case, and the 2 TN cases in which there are two trusted nodes in the network, following corner, diagonal or asymmetric placements.}{2TNGraphsB}

We now consider having two trusted nodes in the network. Specifically, we investigate three methods of placing these two trusted nodes, referred to as {\em corner}, {\em diagonal}, and {\em asymmetric} placements. In the corner placement, the two trusted nodes are located at the two opposite corners of the graph (distinct from Alice and Bob's corners); in the diagonal placement, the two trusted nodes are placed evenly along the diagonal between Alice and Bob; in the asymmetric placement, one trusted node is located in the center, while the other is located one node along the diagonal closer to Alice. Fig.~\ref{fig:TNPlacement} illustrates the above three types of placement for a $7\times7$ grid (we use a $7\times7$ instead of a $5\times5$ grid as the diagonal placement only performs optimally for grid size $N =3x+1$ for some integer $x$).

Figures~\ref{fig:2TNGraphs}(a) through (c) plot the key rate as the decoherence rate increases for each of the above types of placement for the global, IA and NIA routing algorithms, respectively. For comparison, in each figure, the results with a single trusted node placed in the center and no trusted node are also plotted in the figure. We see that for the global routing algorithm (see Fig.~\ref{fig:2TNGraphs}(a)), the diagonal placement outperforms the other four cases (i.e., corner and asymmetric placement of two trusted nodes, as well as no trusted node and placing a single trusted node in the center) by a large margin. 

The clustering of the non-diagonal configurations we attribute to two separate reasons. First, consider that in the central configuration, an optimal pathing results in two channels connecting Alice to the trusted node, and two more connecting the trusted node to Bob. Importantly, these channels will consist of 6 edges each. In the corner configuration, an optimal pathing results in one channel between Alice and each of the two trusted nodes, and likewise from Bob to the trusted nodes. Again, these channels will consist of 6 edges each. As a result, at least in the global case, the likelihood that these paths exist, that the channels are successfully established and that the entanglement remains coherent are all equivalent. It stands to reason then, and the data reflects that reasoning, that the corner configuration and the central configuration are equivalent, in terms of performance. In the other case, it is the asymmetry that causes the equality. Although the second trusted node does result in Alice and the central trusted node sharing a larger key, the central trusted node and Bob establish no more key-bits than if the second trusted node did not exist. As a result, in the final routing of key material, the overall capacity between Alice and Bob remains unchanged, and so the additional key-bits that Alice and the trusted node were able to establish go unused. Finally, the over-performance of the diagonal configuration is a result, as one would expect, from the further segmentation of the $7 \times 7$ network into 3 distinct sub-networks: between Alice and the first trusted node; the first and second trusted nodes; and finally the second trusted node and Bob. As shown in the figure, this additional trusted node can increase our key rate by almost a factor of 1.5 over the central trusted node configuration. 

We repeat the above investigation for the IA algorithm (see Fig.~\ref{fig:2TNGraphs}(b)). Again, we see that the diagonal placement outperforms the others, and that the performance of the corner placement of two trusted nodes is very similar to that when placing a single trusted node in the center. We also see, however, that the redundant trusted node in the asymmetric case results in a large detriment to the key rate, dropping it almost a full .3 key-bits/round. This is likely due to the greedy nature of the routing algorithm -- having an additional node complicates the decision process of the repeaters near both trusted nodes, potentially wasting edges that otherwise could have been utilized in a more effective manner.

The differences in the performance of the different types of placement are even more dramatic for the NIA algorithm in Fig.~\ref{fig:2TNGraphs}(c). For each configuration we see a significant drop in performance, except for the corner placement, which remains relatively consistent with its performance in the other algorithms. This difference can be explained by the fact that that the NIA algorithm does not generally struggle with intersecting paths in the corner case, as the optimal paths themselves tend not to contain many right angles. The effect of this is so great, in fact, that at lower error rates the corner configuration in fact outperforms the diagonal configuration, until the shorter hop-length of the channels in the diagonal configuration becomes the dominating factor in determining key rate. 


As physical distance between Alice and Bob increases, one might think that there is benefit to be had by adding additional repeaters and decreasing the distance between nodes themselves. We investigate whether or not this is feasible in Fig.~\ref{fig:FixedLength} in which the physical distance the network spans is fixed at $10$ km $\times$ $10$ km, while the number of nodes in the grid is increased (this is in contrast to our other simulations where the total width of the network was not fixed, but instead, the length and width are both $(N-1)\cdot L$ km). What we see is that the tendency for BSM failure chance and decoherence probability to increase as path sizes increase puts a damper on any benefit that can be achieved through this method. In some scenarios it is likely that there are trade-offs that can be made, depending on the target distance between Alice and Bob, BSM success probabilities, and decoherence probabilities, but evidently there are limitations inherent to trying to augment the key rate using this method. 

It is clear, however, that this limitation can be alleviated in part by the addition of trusted nodes. As can be seen in Fig.~\ref{fig:FixedLength}, while the key rate of each configuration decreases as the size increases, even as the fiber length decreases, additional trusted nodes remain effective in boosting the key rate. As a result, it is still possible to increase network size to alleviate fiber length concerns, as long as the number of trusted nodes is also increased.  Additionally, 
Fig.~\ref{fig:FixedLength} showcases the way in which non-optimal grid sizes affect key rate for the different TN configurations. Namely, we see that the key rate of the diagonal configuration actually increases between $N=5$ and $N=7$, or $N=9$ and $N=10$, as we move from grid sizes not of the form $3x+1$ to grid sizes of that form. We see the same for the central configuration at the even grid size of $N=10$, where the node is placed slightly off center, and as such actually increases at $N=11$. 

Finally, in Fig.~\ref{fig:2TNGraphsB} we consider the performance of the global and IA algorithms on a number of TN configurations as the BSM success probability increases. Again, we see that the general trend is the same between the global and IA algorithms, with the global algorithm outperforming the IA algorithm in each configuration, and especially the asymmetric case. In the diagonal configuration, with perfect quantum repeaters (i.e. $B=1$), both are able to achieve a key rate of more than .45 key-bits/round with 2 TN in the diagonal configuration, and with 1 TN in the center, the IA algorithm and global algorithms achieve key rates of .3 and .33 key-bits/round, respectively. 
The above results are notable, because they give some sense of the trends we might expect to see as quantum repeater technology advances. They also show the great benefit to using trusted nodes, even with perfect, ideal, repeaters.

\section{Summary and Insights}
The data suggests a number of interesting lessons regarding this sort of QKD network. Most glaringly, it is immediately evident that the addition of even a single trusted node into a network can greatly increase what key rates are achievable, even with ideal repeater technology. Further, we see that a small adjustment to the natural local routing algorithm, in the form of a tendency to avoid intersection of paths, can lead to increases in key rate comparable to, or even exceeding, that of the addition of a single trusted node, with a negligible increase in complexity. As we have seen, this increase in key rates generalizes to multiple trusted nodes, on larger grids, but there is some consideration that must be given to the placement of the additional trusted nodes. Indeed, a non-optimal placement of trusted nodes can in fact hinder the operation of the network, especially for the local algorithms. 

The data further suggests some lessons regarding the limitations of such a network. We see that while fiber length plays an important role in determining key rate, it is the BSM success probability and the decoherence rate that are seemingly the largest obstacles to achieving a higher key rate. This relationship is no more evident than in Fig.~\ref{fig:FixedLength}, where we see that these factors play an important role in limiting the total distance the network can cover, as they make it infeasible to mitigate the effects of fiber length by simply adding additional nodes to the network. In fact  doing so can actually decrease key rates, if not counterbalanced with the addition of trusted nodes. 


Finally, our analysis makes clear the important role trusted nodes will still have to play after quantum repeater networks become practical, or even perfect. As was shown in Fig.~\ref{fig:2TNGraphsB}, even with perfect quantum repeaters, the addition of trusted nodes can result in positive and significant key rates being achieved where they were otherwise not possible. Trusted nodes can be used in the networks to facilitate the establishment of efficient QKD networks reaching along far distances consisting of many nodes. In fact, one can even conceive of network models in which the trusted nodes also operate as part of a multi-party QKD system.  

\section{Conclusion}
In this work we have proposed a novel model for analyzing the performance of quantum repeater QKD grid networks with the inclusion of a minority of trusted nodes. We proposed three routing algorithms, and evaluate the performance of the E91 QKD protocol when using them for a variety of channel and network configurations. We discuss general lessons that can be drawn from our results, including the importance of not only the inclusion of trusted nodes, but also their placement, and some general limitations inherent to working with such networks.  Note that our approach can also be used to determine network resources needed to achieve desired rates between users at given distances or with given repeater quality.  Our work in this paper can  serve as a baseline for future exploration of this area, and we leave open as future work issues such as the development of better local and global algorithms; analysis of mutli-party networks of this form; analysis of more complicated, potentially asymmetric networks; analytic results regarding the capabilities of these networks; the possibility of embedding Alice and Bob as part of a larger network, rather than at the corners; as well as the extension of this analysis to additional QKD protocols. 

\balance
\bibliographystyle{ieeetr}
\bibliography{intro-ref}
\end{document}